\documentclass{emulateapj}
\usepackage{psfig,enumerate}
\usepackage{color}
\usepackage{atbegshi}
\usepackage{url}
\usepackage[caption=false]{subfig}

\newcommand{\dgr}{$^{\circ}~$}

\received{January 5, 2018}
\accepted{July 25, 2018}
\begin{document}

\title{First Data Release of the All-sky NOAO Source Catalog}

\shorttitle{NSC DR1}
\shortauthors{Nidever et al.}

\author{
David L. Nidever\altaffilmark{1,2},
Arjun Dey\altaffilmark{2},
Knut Olsen\altaffilmark{2},
Stephen Ridgway\altaffilmark{2},
Robert Nikutta\altaffilmark{2},
Stephanie Juneau\altaffilmark{2},
Michael Fitzpatrick\altaffilmark{2},
Adam Scott\altaffilmark{2}, and 
Francisco Valdes\altaffilmark{2}
}

\altaffiltext{1}{Department of Physics, Montana State University, P.O. Box 173840, Bozeman, MT 59717-3840 (dnidever@montana.edu)}
\altaffiltext{2}{National Optical Astronomy Observatory, 950 North Cherry Ave, Tucson, AZ 85719}

\begin{abstract}
% From a talk abstract
Most of the sky has been imaged with NOAO's telescopes from both hemispheres. While the large majority of these data were obtained for PI-led projects and almost all of the images are publicly available, only a small fraction have been released to the community via well-calibrated and easily accessible catalogs. We are remedying this by creating a catalog of sources from most of the public data taken on the CTIO-4m+DECam and the KPNO-4m+Mosaic3. This catalog, called the NOAO Source Catalog (NSC), contains over 2.9 billion unique objects, 34 billion individual source measurements, covers $\approx$30,000 square degrees of the sky, has depths of $\approx$23rd magnitude in most broadband filters with $\approx$1--2\% photometric precision, and astrometric accuracy of $\approx$7 mas.
In addition,$\approx$2 billion objects and $\approx$21,000 square degrees of sky have photometry in three or more bands.
The NSC will be useful for exploring stellar streams, dwarf satellite galaxies, QSOs, high-proper motion stars, variable stars and other transients.  The NSC catalog is publicly available via the NOAO Data Lab service.
% photometric precision, proper motions, number of exposures
\end{abstract}


% Outline:
% Introduction
% Observations and Data Reduction
% Results
% Discussion

%NSC DR1 paper
%-abstract
%-intro
%-data set
%-method
%  -measurement
%  -calibration
%  -combination
%-example science use cases
%  -transients, minor planets, planet 9
%  -variable stars
%  -looking for dwarf galaxies
%  -stellar streams
%-future plans
%  -PSF photometry of individual exposures
%  -PSF photometry of coadds
%  -forced photometry?
%  -real-time upgrades
%  -possibly catalog-level transient detection/searches

% Figures
% -map of number of detected objects
% -histogram of seeing
% -histogram of WCS RMS values

% Introduction
\section{Introduction}

It has been clear for more than a decade that the volume of data collected in astronomy is in a period of rapid expansion \citep{Brunner}.  This development is evident in the evolution of area covered by both shallow all-sky surveys and deeper pencil-beam surveys 
(NOAO Deep Wide-Field Survey, \citealt{Jannuzi1999}; Sloan Digital Sky Survey, \citealt{York00}; Pan-STARRS1, PS1 \citealt{Chambers2016}).
%NOAO Deep Wide-Field Survey \citep[NDWFS;][]{Jannuzi1999}, through the Sloan Digital Sky Survey \citep[SDSS;][]{York00},
%to Pan-STARRS1 \citep[PS1;][]{Magnier2016}. 
The trend continues into the future with the Dark Energy Survey \citep[DES;][]{DES}, 
the  Zwicky Transient Factory \citep[ZTF;][]{Bellm2015}, and the
Large Synoptic Survey Telescope \citep[LSST;][]{Ivezic2008}. The growth of Big Data in astronomy can be correctly described as exponential \citep{Zhang} --- a fact that has attracted the attention even of the lay press \citep{Atlantic}. 
These future mega-surveys will define a new astronomical data landscape, with an unprecedented combination of areal coverage, temporal sampling, uniformity, and depth.  

As the community works to evolve astronomy into an increasingly archival and data-intensive science, it faces a conundrum of how to prepare for a research environment that does not yet exist.  Conferences, schools and seminars proliferate. Working groups prepare with multi-year lead times for both narrow and broad science opportunities foreseen for the next decade (e.g., LSST Science Collaborations\footnote{\url{https://www.lsstcorporation.org/node/37}}).  

As we prepare for this fast-approaching era, currently existing and growing data resources offer great potential for immediate analysis, and for exploration and preparatory programs.  A few major archives, such as SDSS\footnote{\url{http://www.sdss.org}}, play an essential role, offering opportunities that can
%put in service of helping to train the astronomical community for the fast-approaching LSST era
be studied with the most modern data mining and computer learning methodology, developing expertise, experience, and crucially, producing science results now --- an essential ingredient for a healthy science demographic. 
% global sentence of paragraph
% SDSS have provided these tools to mine these data
% these archives offer opportunities for exploration, new methodologies, etc.

\begin{figure*}[ht]
\begin{center}
\includegraphics[trim={0cm 4.9cm 2cm 1cm},clip,width=1.0\hsize,angle=0]{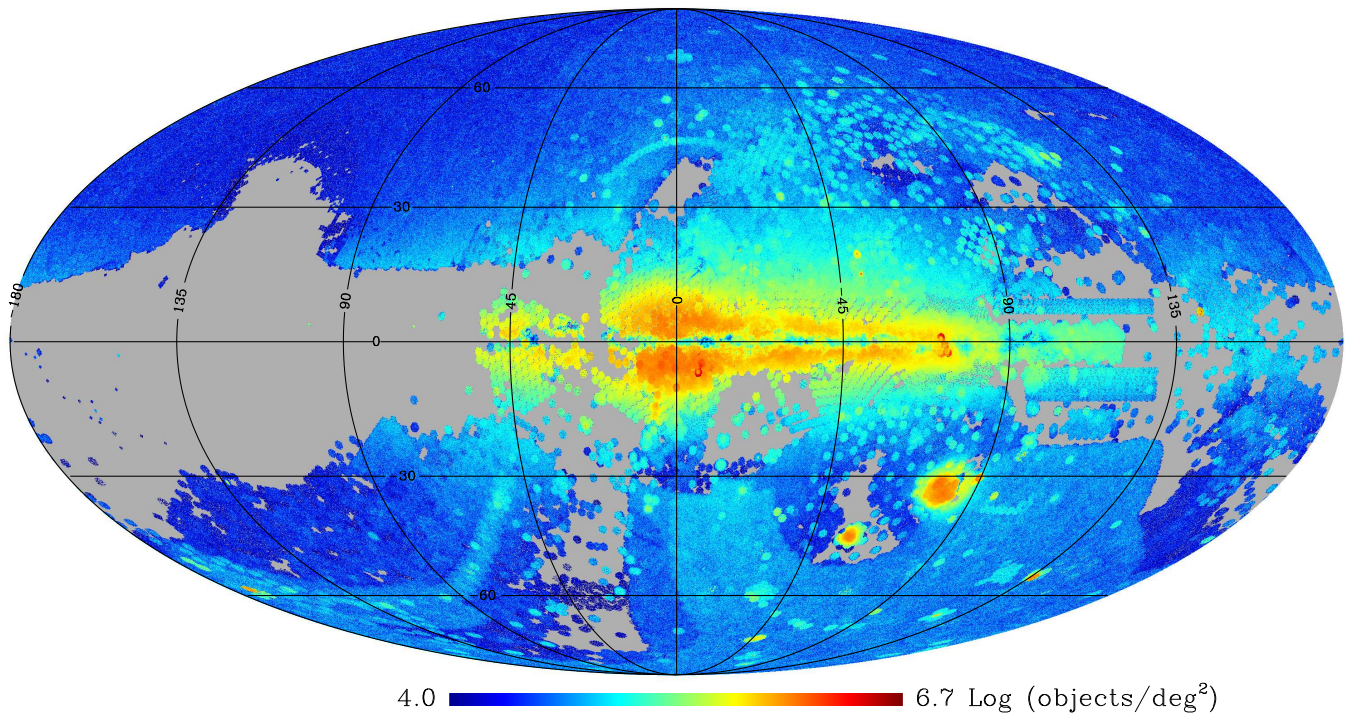}
\end{center}
\caption{Density of the 2.9 billion NSC objects on the sky in Galactic coordinates.  The higher densities from the Galactic midplane and Bulge as well as the LMC and SMC are readily apparent.  The density is a combination of the true density of objects as well as the particular exposure times of the various observing programs.}
\label{fig_bigmap}
\end{figure*}

It is less well known that the major observing facilities on Kitt Peak and Cerro Tololo have, over the years, delivered high quality imaging data sets with broad areal and wavelength coverage. Though originally intended for a diverse variety of science programs, these data have a considerable degree of homogeneity. Since the bulk of these data were obtained under different protocols and by multiple PI's, they are not conveniently accessed and simply merged. Recognizing the potential inherent in these massive data, we have undertaken to convert them into a unified, public database.  By uniformly photometering the reduced images,
%By reprocessing the raw data,
with consistent quality control, selection, calibration and documentation, we have merged these data as the NOAO Source Catalog (NSC). Reductions have utilized proven tools and methodologies, adapted to the extant data materials. Careful attention has been paid to crowding, seeing and image shape, photometric calibration and astrometric zero points.

The NSC is now ``open for business", with over 255,000 exposures and 2.9 billion unique objects (Figure \ref{fig_bigmap}). The NSC may be accessed through NOAO's Data Lab, with tools for discovery, image cutout, virtual storage, and supported Jupyter notebook analysis, and also by direct query or TAP service.  With a recognized trajectory for extensions and improvements, the NSC will be updated continuously as new material is added, as more advanced processing can be implemented, and as the database is further adapted to user needs.  The layout of this paper is as follows. Section \ref{sec:dataset} discusses the imaging dataset while Section \ref{sec:methods} gives details on the reduction and photometric methods.  The final catalog and achieved performance are discussed in Section \ref{sec:dr1} and several science use cases are presented in 
Section \ref{sec:usecases}.  Future plans for improvement are detailed in Section \ref{sec:future} and a brief summary is given in Section \ref{sec:summary}.

% one could point our that we can readily notice the higher object density along the galactic plane, and Galactic Bulge, as well as at the location of the SMC and LMC. 
%It could also be mentioned that the number density of objects is a combination of both the true number density and the depth of the exposures? (and that some streaks/small overdensities come from deeper PI
% projects)

\section{Dataset}
\label{sec:dataset}

All data presented in this paper are drawn from the NOAO Science Archive.  While the Archive contains data sets from all telescopes and instruments, the large majority of the NSC data are from the Dark Energy Camera (DECam) on the CTIO-4m with over 195,000 exposures used in NSC DR1.  The Archive also contains a large volume of public data for the DESI Legacy Surveys\footnote{\url{http://legacysurvey.org}} (Dey et al.\ 2018, in preparation) including $z$-band images from KPNO-4m Mayall + Mosaic3 (Mayall $z$-band Legacy Survey; MzLS; \citealt{Dey2016}) and $g$ and $r$-band image from Bok-2.3m+90Prime (Beijing-Arizona Sky Survey; BASS; \citealt{Zou2017}).  The MzLS and BASS data cover a good fraction of the northern sky and were also included in NSC DR1.  All three cameras have multiple CCD detectors -- DECam has 62\footnote{One DECam CCD stopped functioning shortly after commissioning, one is only partially usable, and one was inoperative for several years.} while both Mosaic3 and 90Prime have four.
Table \ref{table_surveys} shows the top 11 contributers of data to the NSC. Together, these programs constitute 71\% of the total NSC DR1 holdings.

\begin{center}
\begin{deluxetable*}{lllcl}
\tablecaption{NSC Surveys}
\tablecolumns{5}
\tablehead{
  \colhead{Name} & \colhead{PI} & \colhead{PropID} &\colhead{N$_{\rm exposures}$} & \colhead{Bands} 
}
\startdata
DES\tablenotemark{a} & Friedman & 2012B-0001 & 58,546 & $grizY$ \\
Legacy Surveys:\tablenotemark{b} \\
\ \ \ DECaLS\tablenotemark{c}  & Schlegel \& Dey & 2014B-0404 & 17,533 & $grz$ \\
\ \ \ BASS\tablenotemark{d} & Zou \& Fan & 2015A-0801 & 19,741 & $gr$ \\
\ \ \ MzLS\tablenotemark{e} & Dey & 2016A-0453 & 40,224 & $z$ \\
NEO\tablenotemark{f} & Allen & 2013A-0724 / 2013B-0536 & 11,800 & {\em VR} \\
DECaPS\tablenotemark{g} & Finkbeiner & 2016A-0327 / 2016B-0279 & 7,590 & $grizY$ \\
Bulge Surveys:\\ 
\ \ \ BDBS\tablenotemark{h} & Rich & 2013A-0529 / 2014A-0480 & 3,849 & $ugrizY$ \\
\ \ \ DSSGB\tablenotemark{i} & Saha & 2013A-0719 & 3,837 & $ugriz$ \\
Light Echoes\tablenotemark{j} & Rest & 2013A-0327 / 2014A-0327 / 2014B-0375 / 2015A-0371 / 2016A-0189 & 7,622 & $r${\em VR} \\
SMASH\tablenotemark{k} & Nidever & 2013A-0411 / 2013B-0440 & 6,645 & $ugriz$ \\
BLISS\tablenotemark{l} & Soares-Santos & 2017A-0260 & 3,049 & $griz$
\enddata
\label{table_surveys}
\footnotetext{Dark Energy Survey \citep{DES}; \url{https://www.darkenergysurvey.org}}
\footnotetext{\url{http://legacysurvey.org}}
\footnotetext{DECam Legacy Survey; \url{http://legacysurvey.org/decamls/}}
\footnotetext{(Beijing-Arizona Sky Survey \citep{Zou2017}}
\footnotetext{Mayall $z$-band Legacy Survey \citep{Dey2016}}
\footnotetext{DECam Near Earth Object survey \citep{Allen2016}}
\footnotetext{DECam Plane Survey \citep{Schlafly2018}; \url{http://decaps.skymaps.info}}
\footnotetext{Blanco DECam Galactic Bulge Survey \citep[BDBS;][]{Rich2015}}
\footnotetext{Deep Synoptic Study of the Galactic Bulge}
\footnotetext{Light Echoes of Galactic Explosions \citep{Rest2015}}
\footnotetext{Survey of the Magellanic Stellar History \citep{Nidever2017}; \url{http://datalab.noao.edu/smash/smash.php}}
\footnotetext{Blanco Imaging of the Southern Sky}
\end{deluxetable*}
\end{center}

% * <adam.c.scott@gmail.com> 2017-12-29T16:41:46.790Z:
% 
% I'll take a crack at how much of the fraction:
% 
% Area of sphere in square degrees is 4pi * (180/pi)^2 ( or ~ 41252.96124941927103129467). Half of that (northern sky) would be 20626.48062470963551564734 square degrees.    I believe BASS says it covers ~ 5400 square degrees (http://batc.bao.ac.cn/BASS/doku.php?id=datarelease:survey_footprint:home) which makes 26.17% of northern sky (5400/20626.48).  How's that? :)
%  
% 
% ^ <adam.c.scott@gmail.com> 2017-12-29T17:17:57.424Z.
% could mention FOV, nchips and pixel scale for all three imagers.

Older images (primarily from the Mosaic-1, 1.1 and 2 cameras) have not been added to the NSC yet mainly because of the difficulty of photometrically calibrating so many different filters.  We are investigating means of including these in future versions of the NSC.

\section{Reduction and Photometry}
\label{sec:methods}

%Mention that we use the CP InstCal images.

The NSC data processing makes use of three separate software packages: (1) the
NOAO Community Pipelines for instrumental calibration (\citealt{Valdes2014}; Valdes et al., in preparation)\footnote{\url{https://www.noao.edu/noao/staff/fvaldes/CPDocPrelim}},
%technical descriptions are in progress; for DECam an early description is \cite{Valdes2014}
%and a more technical draft is here\footnote{\url{https://www.noao.edu/noao/staff/fvaldes/CPDocPrelim}}),
(2) Source Extractor\footnote{\url{https://www.astromatic.net/software/sextractor}}
\citep{Bertin1996} for performing source detection, photometry, and parameter estimation from the images, and (3)
custom software\footnote{\url{https://github.com/dnidever/noaosourcecatalog}} (in python and IDL)
to run SExtractor on the images and to perform photometric and astrometric calibration and combination.

\subsection{Community Pipeline Calibrations}
\label{subsec:reduction}

All of the NSC images were instrumentally calibrated by the NOAO Community Pipelines (``CPs'').
The CPs are NOAO pipelines for each instrument which produce the data products in the NOAO archive. The CPs include algorithms and code (from a variety of sources) which are modified and packaged for the needs of the NOAO environment and characteristics of the different instruments.  The environment includes a common orchestration framework for distributed processing and common data formats in the NOAO archive.

\begin{center}
\begin{figure*}[t]
$\begin{array}{cc}
\includegraphics[width=0.5\hsize,angle=0]{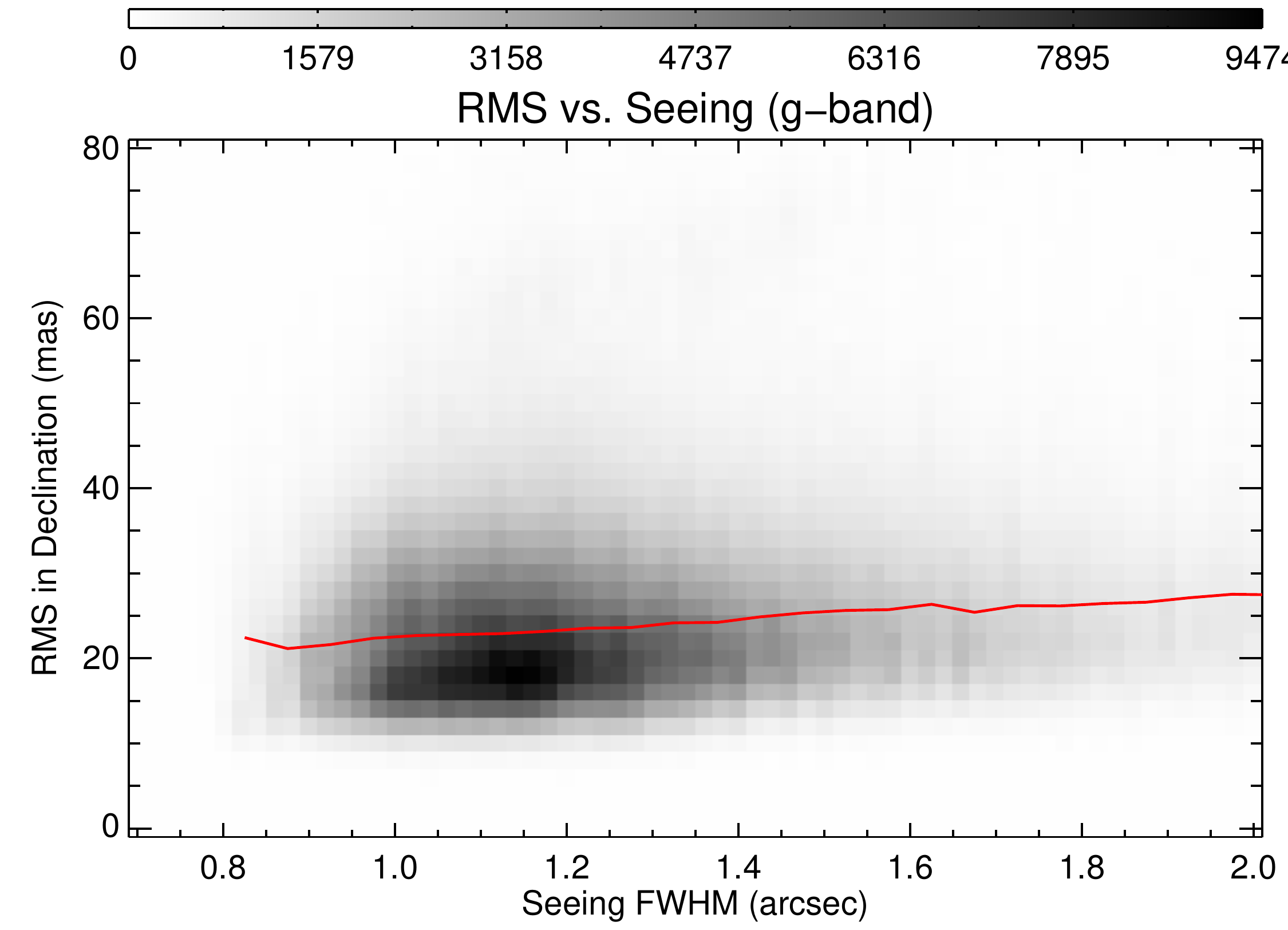}
\includegraphics[width=0.5\hsize,angle=0]{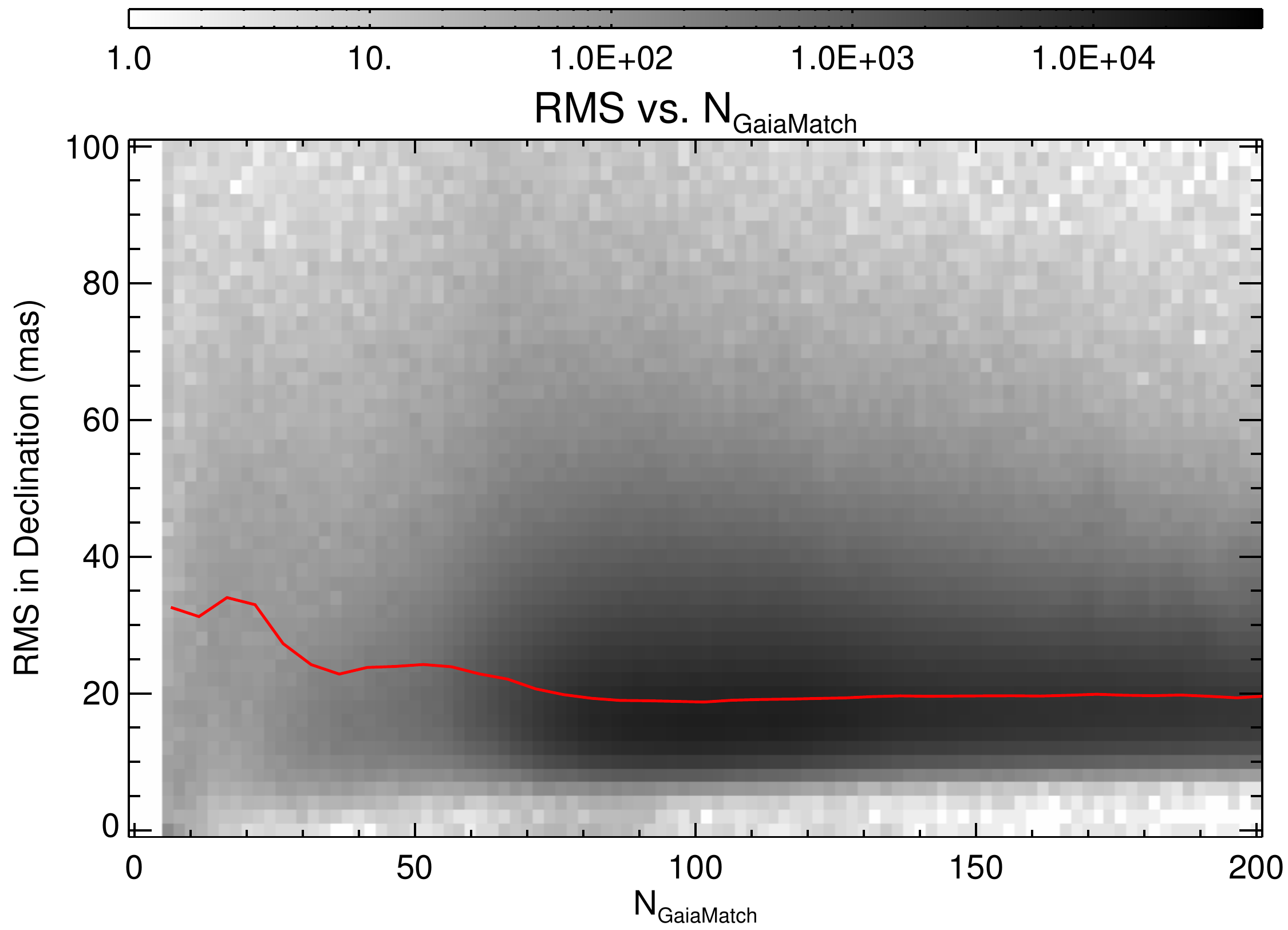}
\end{array}$
\caption{The RMS scatter in $\delta$ of CCD astrometric fits using Gaia DR1 data.  (Left) RMS versus seeing (for $g$-band and airmass$<$1.5) on a linear scale, and (right) RMS versus the number of Gaia matches on a logarithmic scale (for seeing$<$1.5\arcsec~and airmass$<$1.5).  Binned median values are shown by the red solid curves.}
\label{fig_decrms}
\end{figure*}
\end{center}

The CPs create several types of images: (a) ``InstCal" -- instrumentally calibrated images for each exposure, (b) ``Resampled'' -- Instcal images resampled onto the common tangent plane projection for exposures of this field; and (c) ``Stacked'' -- coadded Resampled images of the same filter, field and similar exposure time.  We use the Instcal images (hereafter ``CP images'') for further analysis to create the NSC catalogs.   There are separate pipelines for each camera (DECam, Mosaic3, 90Prime) that share similar steps and software modules though with variations for different characteristics of each camera.  The main calibration operations for the CP images are as follows.
\begin{itemize}
\item A linearity correction as a function of count levels is applied for DECam only.
\item The crosstalk signal between amplifiers is subtracted using empirically derived amplitude coefficients.
\item The electronic bias is estimated from overscan regions and subtracted.
\item The nightly zero level bias image correction and dome flat fielding from master dome calibrations. Dark current is negligible in all three cameras and no dark count subtraction is applied.
\item The amplifier gain across all CCDs in an image is ``balanced'' based on a photometric comparisons of detected sources.  For DECam this is determined from widely dithered calibration exposures called ``star flats'' \citep{Bernstein2017b}.  For Mosaic3 and 90Prime this is done by reference to PS1 photometry.  For Mosaic3 this is done as a function of the sky level.
\item Non-linear, saturated, or bleed pixels are identified, masked and interpolated. The cores of saturated stars are not replaced by interpolation.
\item Shifted lines and columns in the Mosaic3 and 90Prime images are corrected.
\item The images are astrometrically calibrated and the image World Coordinate System (WCS) is recorded in the headers. For DECam this uses SCAMP \citep{Bertin2006} to fit a constrained polynomial distortion model with 1st and 2nd order adjustments referenced to 2MASS or, more recently, Gaia DR1.  For Mosaic3 and 90Prime this uses IRAF/ccmap with a fixed radial distortion model and 4th order polynomial perturbations referenced to Gaia DR1.  The calibrated WCS is TPV\footnote{{\url https://fits.gsfc.nasa.gov/registry/tpvwcs/tpv.html}}.
\item Cosmic-rays and other particle events are identified and masked. For Mosaic3 and 90Prime data we use an NOAO algorithm to identify particle events and interpolate over them; for DECam, we use an LSST algorithm and do not interpolate \citep[the thick CCDs in Mosaic3 and DECam produce many particle events][]{Groom2004}.  Streak detection and masking is done similarly for all three cameras using an NOAO algorithm.
\item Sky patterns and gradients are removed.
\item ``Pupil ghosts" are removed for all DECam filters and for $g$ and $z$ in the other two cameras.  For DECam the pupil is estimated for each exposure
by azimuthal fitting. For Mosaic3 and 90Prime a template is extracted from unregistered stacks and then scaled and subtracted from
each exposure.
\item A fringe pattern is subtracted for DECam $z$ and $Y$, Mosaic3 $z$, and 90Prime $r$ exposures.  For DECam a manually derived fringe template is used and for the other cameras a template is derived from a dark sky stack.  Note that DECam fringe removal only began in 2018.
\item An illumination flat field for DECam is applied using a ``dark sky illumination'' template created from long exposures in each band over several nights.
\item Special filtering is used to suppress a fairly strong variable pattern noise in Mosaic3 images and a variable stripping pattern in 90Prime images.
\end{itemize}

\subsection{Measurement}
\label{subsec:measurement}

We considered several software packages for performing the photometric source extraction. 
Source Extractor (SExtractor) was chosen because it is well tested, produces source morphology parameters (which are useful for galaxy science), is fast, and can give useful results in crowded regions if tuned properly.

SExtractor was run on each individual CP-processed CCD ``flux'' image to extract aperture photometry and morphological parameters.  The CP inverse variance images were used as weight images (i.e., the SExtractor {\tt WEIGHT\_IMAGE}) and the weights were set to zero for pixels masked as bad (non-zero) in the CP ``quality mask'' image.  A {\tt WEIGHT\_THRESH}=1$\times$10$^{-8}$ was used to indicate to SExtractor to ignore the bad pixels.  The CP quality mask image was also used for the {\tt FLAG\_IMAGE} to give mask information for individual sources.
The CP mask flags were converted to bitmasks (see Table \ref{table_cpflags}) and homogenized across the various CP versions and are similar to the Pre-V3.5.0 CP flags.  The V3.5.0 CP values are integers and are not properly interpreted by SExtractor especially when combining values across pixels in a source footprint.
The SExtractor configuration parameters\footnote{\url{https://github.com/dnidever/noaosourcecatalog/blob/master/params/default.config}} were modified slightly for each image based on the CCD parameters (gain, saturation level and pixel scale, which in turn affect the aperture sizes) and weather conditions (seeing FWHM).

\begin{center}
\begin{deluxetable}{lcl}
\tablecaption{CP Bitmask Flags}
\tablecolumns{3}
\tablehead{
  \colhead{Bit} & \colhead{Value} & \colhead{Condition} 
}
\startdata
1 & 1 & Bad pixel \\
2 & 2 & not used \\
3 & 4 & Saturated \\
4 & 8 & Bleed mask \\
5 & 16 & Cosmic ray \\
6 & 32 & Low weight \\
7 & 64 & Diff detect
\enddata
\label{table_cpflags}
\end{deluxetable}
\end{center}

One of the goals of the NSC (as for most surveys) is to achieve uniformity across the survey in the detection limit (e.g., 5$\sigma$) for all filters and seeing conditions and to also perform well in crowded and uncrowded regions.  Accomplishing both goals simultaneously is challenging.  The best detection is achieved when the image is smoothed with a kernel the size of the PSF \citep{Bijaoui1970}.  However, for poor seeing exposures in crowded regions this smoothing can cause a very large fraction of the sources in the image to be blended together and produces difficulties for the deblending algorithm.  As a compromise, we decided to use a consistent smoothing kernel (FWHM=4 pixels) and detection threshold per pixel (1.1$\times$ the background noise) for all exposures.  This results in a $\approx$5$\sigma$ source detection limit for the median seeing ($\approx$1.26\arcsec) images (for DECam and Mosaic3) and does not overwhelm the deblending algorithm in the most crowded regions.  Variations on this scheme will be evaluated for future versions of the NSC.
% * <adam.c.scott@gmail.com> 2017-12-29T17:23:30.391Z:
% 
% For me it makes sense to punctuate like this: "..for poor-seeing exposures in crowded regions, a very large fraction..."
% 
% ^ <adam.c.scott@gmail.com> 2017-12-29T17:25:05.638Z.

Smoothing (``filtering") an image in SExtractor results in bad pixels affecting their neighbors. 
To compensate, we artificially grew the bad pixel masks around these regions in the mask (``flag'') image using a 7$\times$7 kernel (same size as our filtering kernel); this enabled us to flag the affected sources and ignore them in the catalog.
% 7 pixels wide filter convolution kernel

\begin{center}
\begin{figure}[t]
\includegraphics[width=1.0\hsize,angle=0]{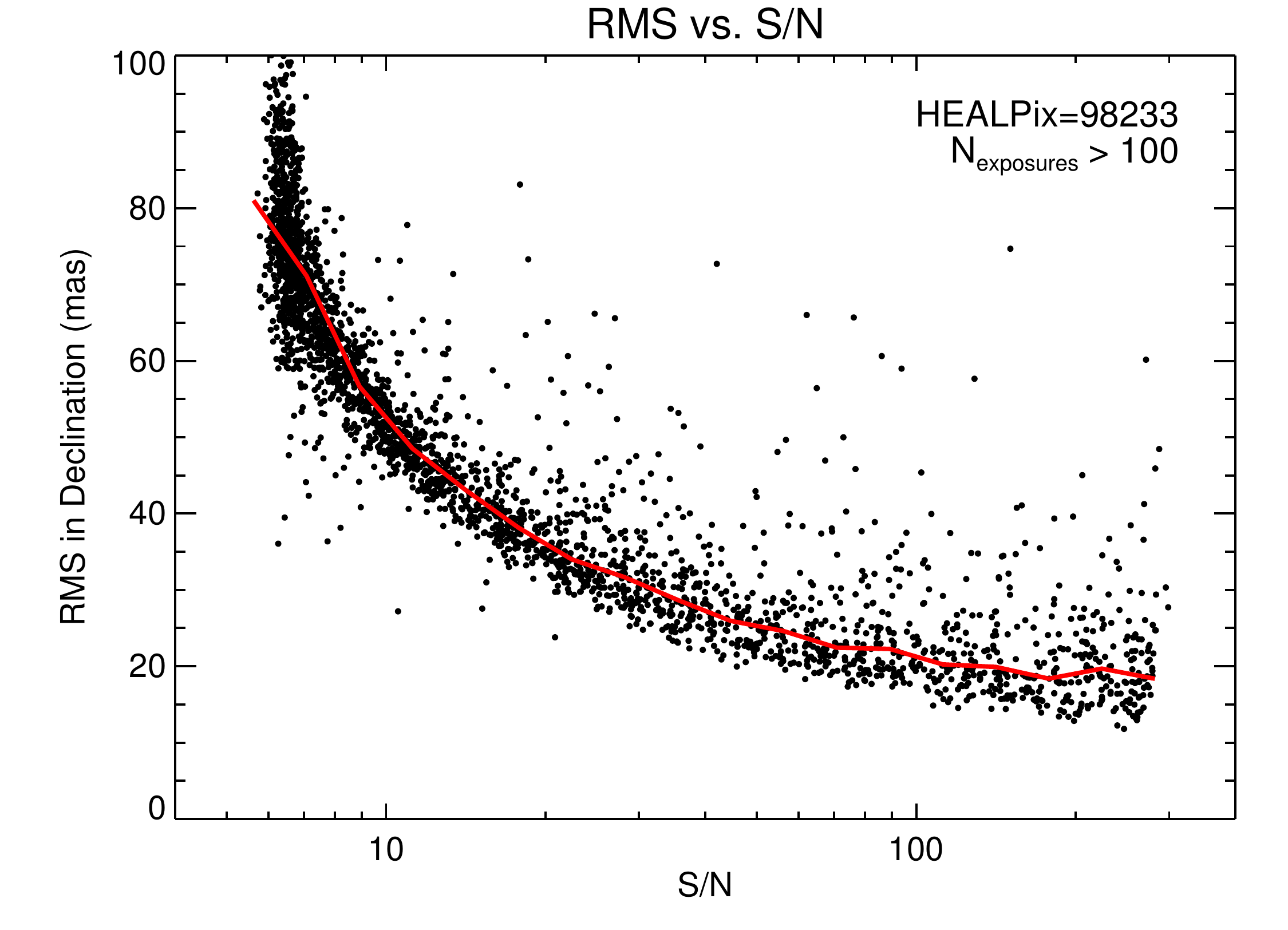}
\caption{The RMS scatter in $\delta$ across multiple independent detections of stars versus S/N (for stars with N$_{\rm exposures}$$>$100 and seeing$<$1.5\arcsec). Binned median values are shown by the red solid curve.  The mean RMS for high-S/N stars is $\sim$20 mas.}
\label{fig_decrms_snr}
\end{figure}
\end{center}

We use the SExtractor automatic aperture magnitudes ({\tt MAG\_AUTO}, which estimates the magnitude in an elliptical aperture defined by a source's morphology) as the primary photometric measurement.
Aperture photometry for multiple apertures (1\arcsec, 2\arcsec, 4\arcsec~and 8\arcsec~diameters) were also computed and are provided in the individual measurements table (see Section \ref{sec:dr1}).

\subsection{Calibration}
\label{subsec:calibration}

\subsubsection{Astrometry}
\label{subsubsec:astrometry}

Although the CP performs astrometric calibration of each image, we found that these solutions sometimes had errors of a few tenths of an arcsecond.
Therefore, we used Gaia DR1 data to apply additional, small-scale, linear astrometric correction terms (on top of the CP solution) for each CCD independently.
Crossmatches were found between well-measured NSC sources (good astrometric and photometric measurements and no bad quality flags) and Gaia DR1 sources using a 0.5\arcsec~matching radius, The coordinates of both groups were then transformed to a tangent plane projection using the center of the chip ({\tt RA} and {\tt DEC} in the {\tt chip} table) giving $\zeta$ ($\alpha$ direction) and $\eta$ ($\delta$ direction).  Linear additive correction coefficients in both $\zeta$ and $\eta$ (tangent plane longitude and latitude coordinates) were then determined (with proper weighting using the NSC and Gaia astrometric uncertainties) for both $\alpha$ and $\delta$:

$\alpha$ corr = {\tt RA\_COEF1} + {\tt RA\_COEF2}$\times$$\zeta$ + {\tt RA\_COEF3}$\times$$\zeta\eta$ + {\tt RA\_COEF4}$\times$$\eta$ \\
$\delta$ corr = {\tt DEC\_COEF1} + {\tt DEC\_COEF2}$\times$$\zeta$ + {\tt DEC\_COEF3}$\times$$\zeta\eta$ + {\tt DEC\_COEF4}$\times$$\eta$ \\

\noindent
There are roughly 200 Gaia DR1 matches per chip that are used for astrometric calibration. The derived best-fit coefficients are released as part of the DR1 tables (see Section \ref{sec:dr1}).

A robust standard deviation (using the median absolute deviation) and an associated standard deviation of the mean are calculated for the residuals of the linear astrometric fits (for point sources with S/N$>$50) in $\alpha$ ({\tt RARMS, RASTDERR}) and $\delta$ ({\tt DECRMS, DECSTDERR}).  The median RMS value in $\alpha$/$\delta$ is 22 mas and the median standard deviation of the mean is 1.9 mas.  The RMS varies by $\sim$10\% across the various filters and is $\sim$20\% larger for 90Prime than the other two cameras.  There is also an increase in RMS with seeing from 20 mas at 0.8\arcsec~to 28 mas at 2.0\arcsec~(Figure \ref{fig_decrms}).  In addition, there is a dependence of the scatter on the number of Gaia matches (below 70 matches) increasing to 35 mas at 5 matches (Figure \ref{fig_decrms}).  The scatter also degrades for very short exposure times ($<$10s) increasing to 55 mas at $\sim$1s.
From regions with hundreds of exposures, the astrometric scatter over multiple independent measurements for high S/N stars (S/N$>$150; with random astrometric uncertainties $\sim$5 mas) is $\sim$20 mas (Figure \ref{fig_decrms_snr}).  This suggests that any systematics in the astrometry are roughly at this level which is consistent with the known astrometric systematics in DECam data including imperfections in the CCD substrate (``tree rings") and differential chromatic aberration \citep{Bernstein2017a}.
However, the systematics can be reduced by averaging over multiple independent measurements.  For high S/N stars with many detections the RMS scatter of their {\em average} positions compared to Gaia DR1 is $\sim$7 mas.

\subsubsection{Photometry}
\label{subsubsec:photometry}

% Images of the model photometry.

%For these model magnitudes, we used the all-sky catalogs 2MASS (Skrutskie et al. 2006), Gaia DR1 (Gaia Collaboration et al. 2016), APASS (Henden et al. 2015), and GALEX (Bianchi et al. 2011), along with the Schlegel, Finkbeiner & Davis (1998, SFD) maps for Galactic reddening, to construct linear combinations of measurements that best approximated PS1 grizY photometry, u-band photometry from the SMASH survey (Nidever et al. 2017), and Gaia G (as a surrogate for VR) over most of the range in color. This technique is similar to that used for the “real-time calibration” performed by Pan-STARRS1 (PS1; Chambers et al. 2016).

For large and diverse datasets like the NSC, it is challenging to use the classical calibration method of 
standard star fields to derive photometric transformation equations because this is often a manual and time-consuming process.  Instead, it is more common to derive zero points for each exposure (or CCD image) by comparing to existing photometric catalogs in the same area of the sky and in the same filters. 
% ascott: I found "Objects matched to the 2MASS catalog (at least 100 in each field) were used to determine a zero point for each filter in each deep image" as an example in this survey http://iopscience.iop.org/article/10.1088/0004-6256/138/2/402/pdf  Is there a better example I can find?
Since there is currently no existing large-scale, public survey in optical broad-band filters in the southern hemisphere, this was only an option for some bands in the northern portion of our data.  We used the PS1 photometry to calibrate all $grizY$ data north of declination {\bf $\delta=-29^\circ$} which includes all of the data from the Mosaic-3 and 90Prime cameras.  For all data south of {\bf $\delta=-29^\circ$}, and for the $u$ and {\em VR} bands all over the sky, we instead constructed model magnitudes to perform the photometric calibration. For these model magnitudes, we used the all-sky catalogs 2MASS \citep{Skrutskie2006}, Gaia DR1 \citep{GaiaDR1}, APASS \citep{Henden2015}, and GALEX \citep{Bianchi2011}, along with the  \citet[][SFD]{Schlegel1998} maps for Galactic reddening, to construct linear combinations of measurements that best approximated PS1 $grizY$ photometry, $u$-band photometry from the SMASH survey \citep{Nidever2017}, and Gaia $G$ (as a surrogate for {\em VR}) over most of the range in color. This technique is similar to that used for the PS1 ``real-time calibration" \citep{Magnier2016}.

The model magnitude coefficients were determined in two steps:

\begin{center}
\begin{figure*}[t]
$\begin{array}{cc}
%trim={0cm 1.2cm 0.8cm 3cm},clip,
\includegraphics[width=0.5\hsize,angle=0]{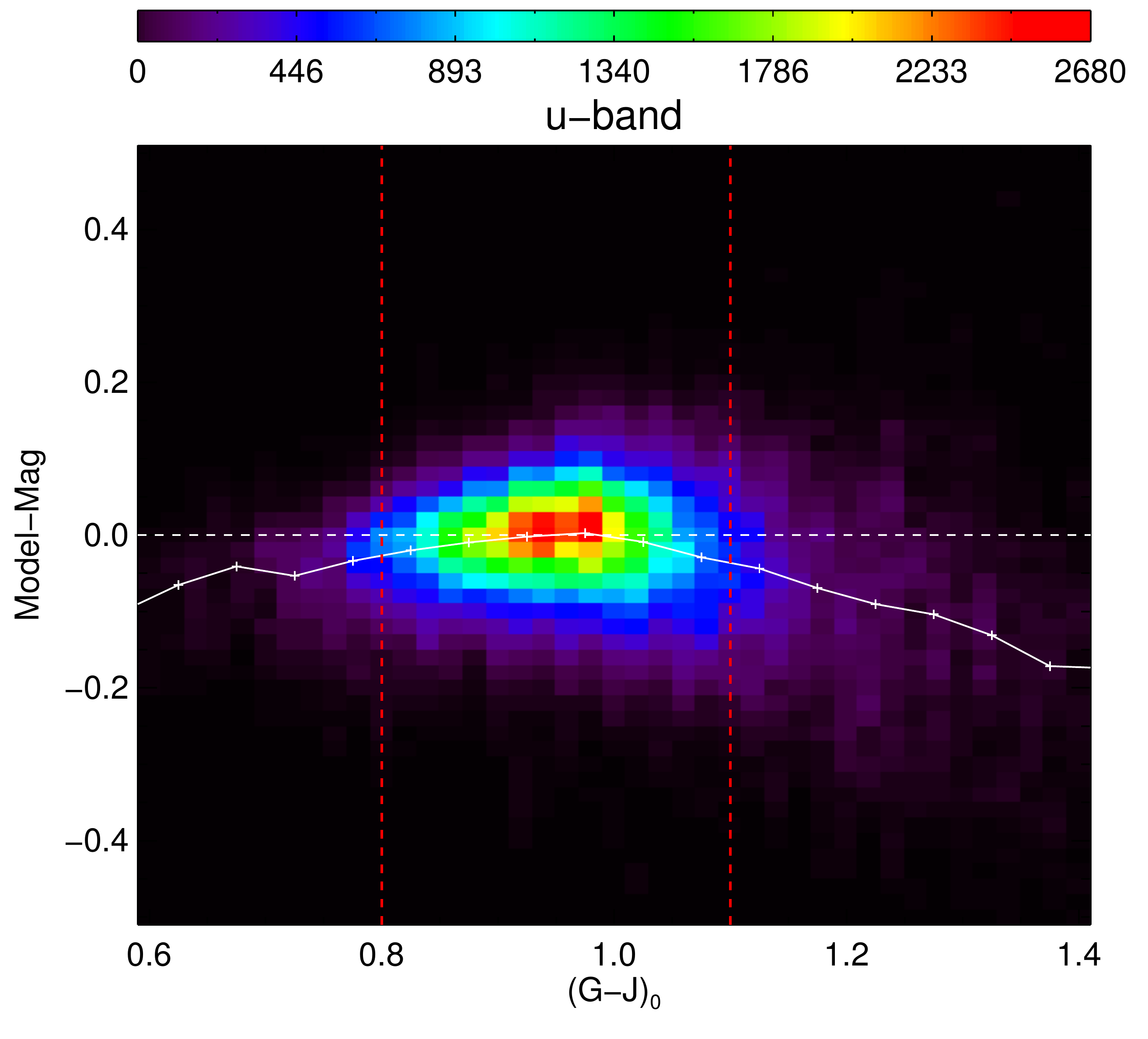}
\includegraphics[width=0.5\hsize,angle=0]{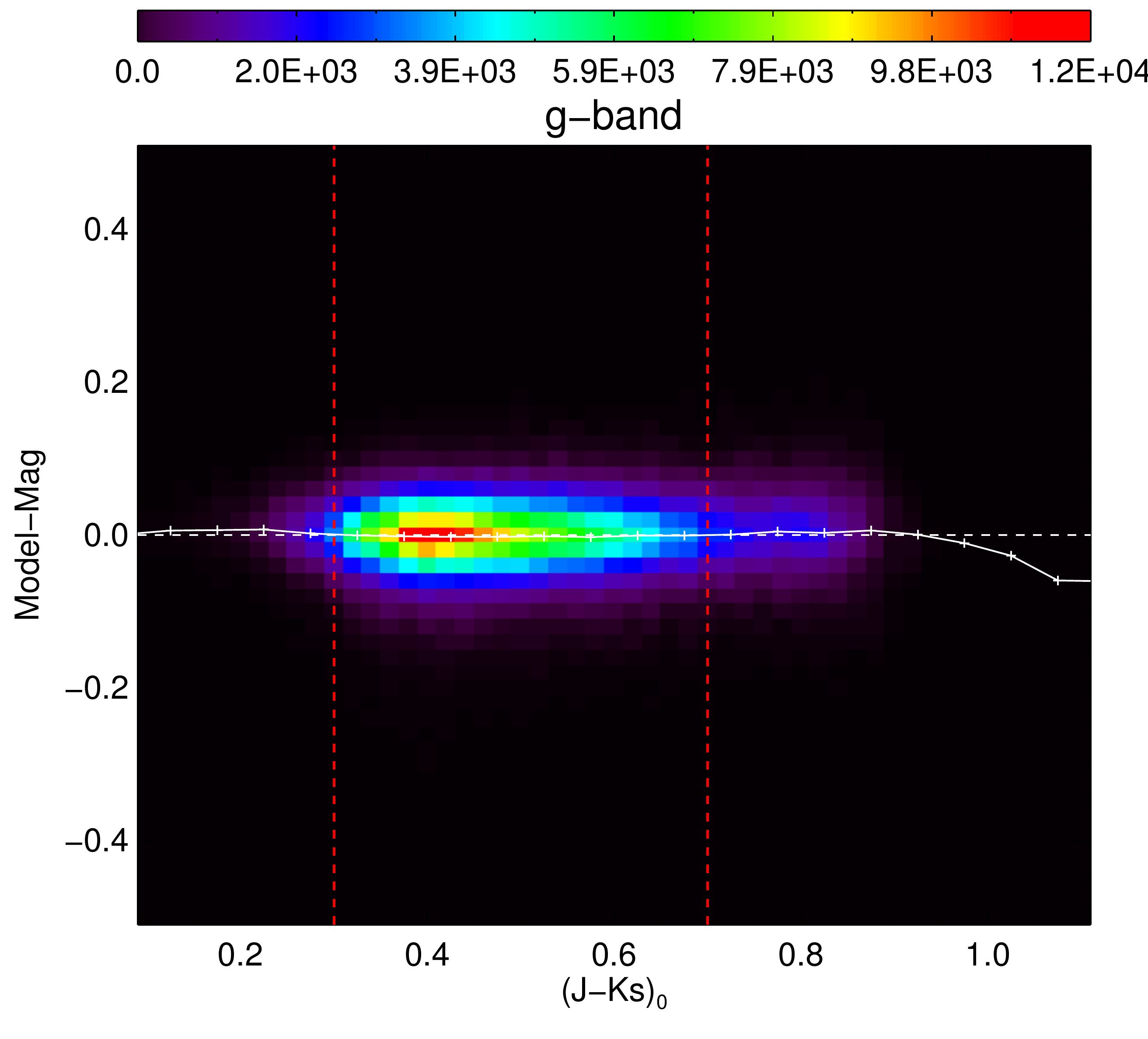} \\
\includegraphics[width=0.5\hsize,angle=0]{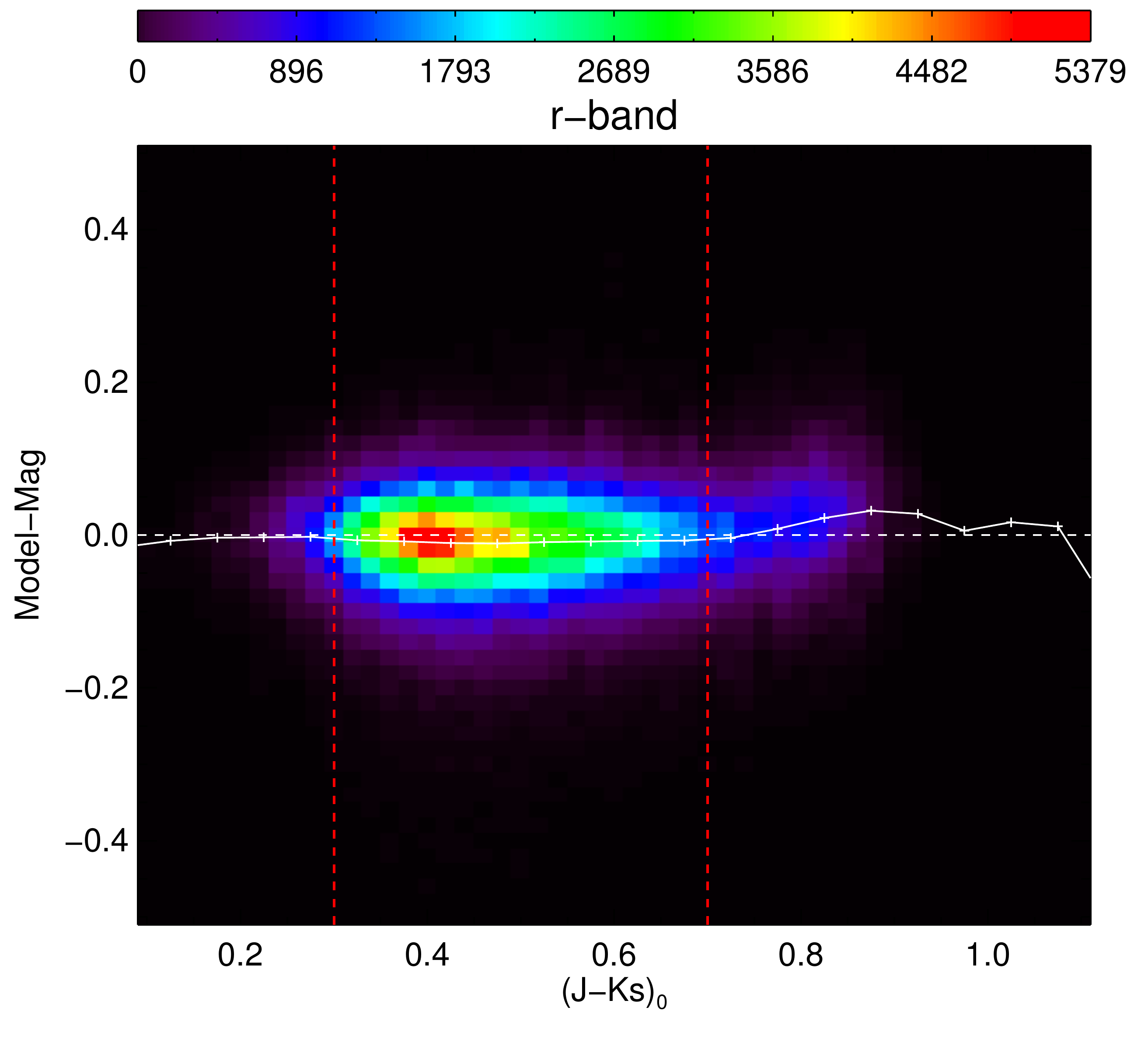}
\includegraphics[width=0.5\hsize,angle=0]{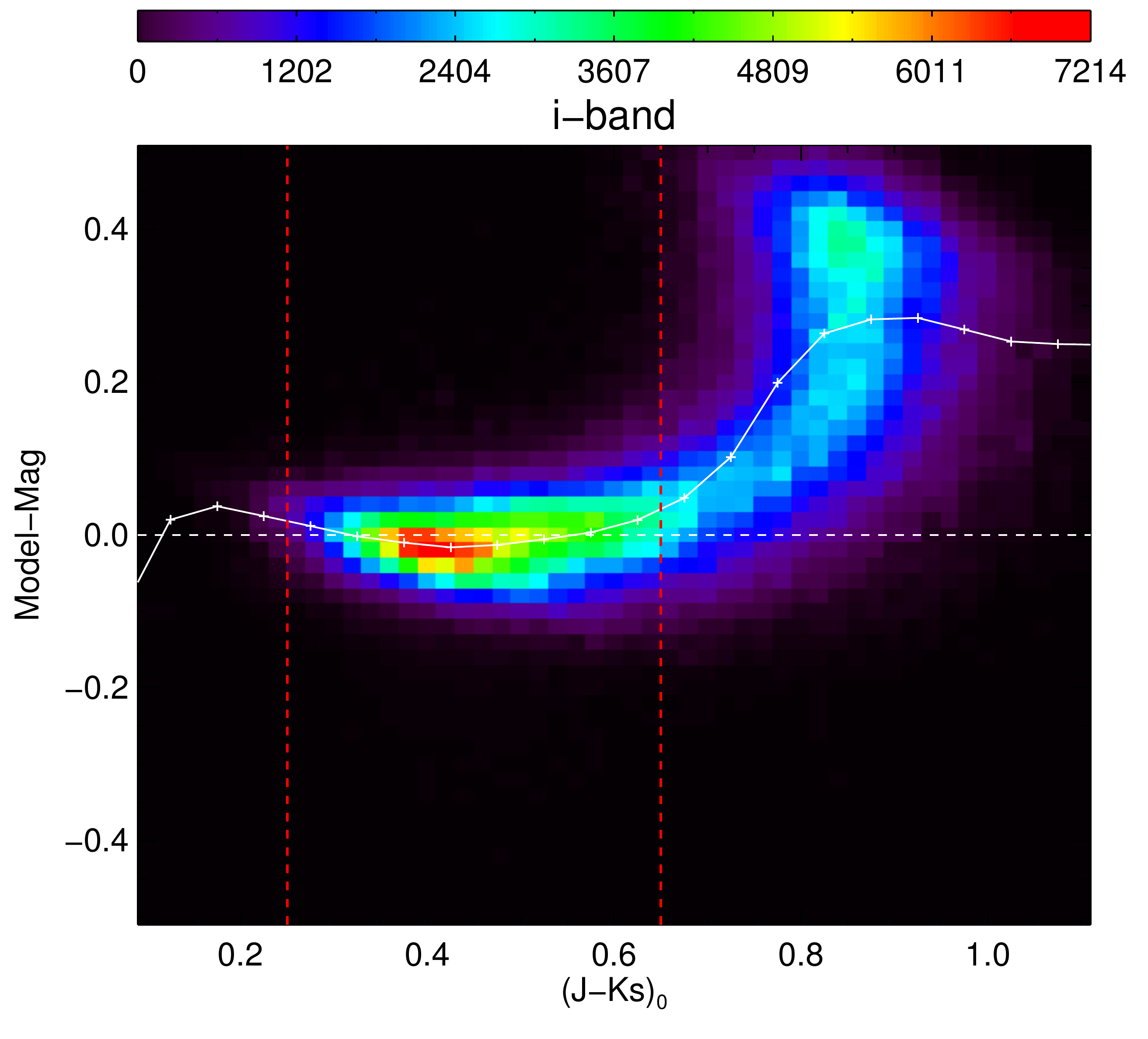}
\end{array}$
\caption{Model magnitude figures showing the density of stars in residual magnitude (model minus reference magnitude) versus color
with a separate panel for each band. The median residual values in color bins are shown as white crosses and connected with solid white lines. 
The color ranges used to determine the zero-point per exposure are indicated by the two vertical red dashed lines.
Note that the colors and color ranges shown change from panel to panel.}
\label{fig_modelmag}
\end{figure*}

\begin{figure*}[ht!]
\ContinuedFloat
$\begin{array}{cc}
\includegraphics[width=0.5\hsize,angle=0]{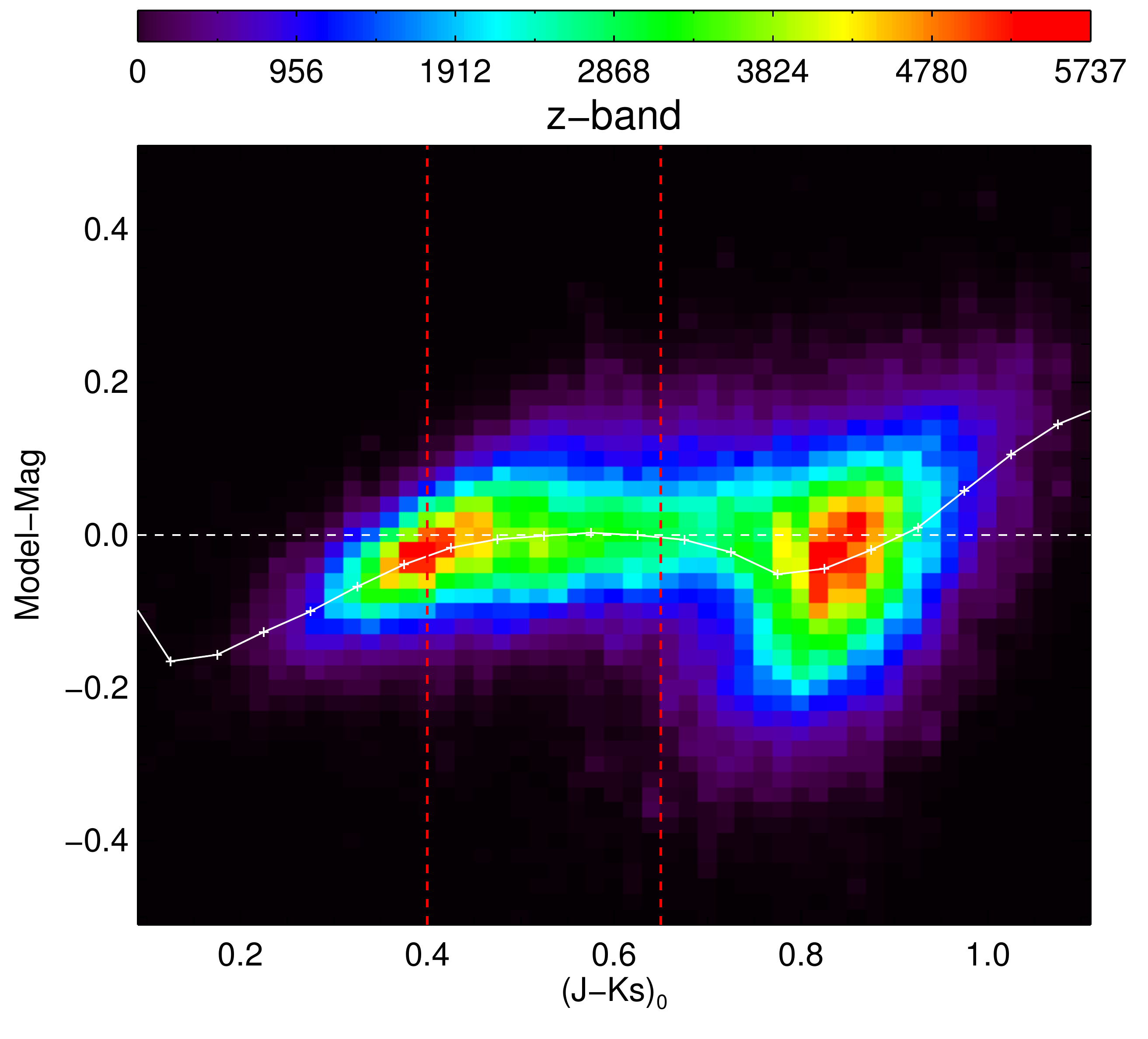}
\includegraphics[width=0.5\hsize,angle=0]{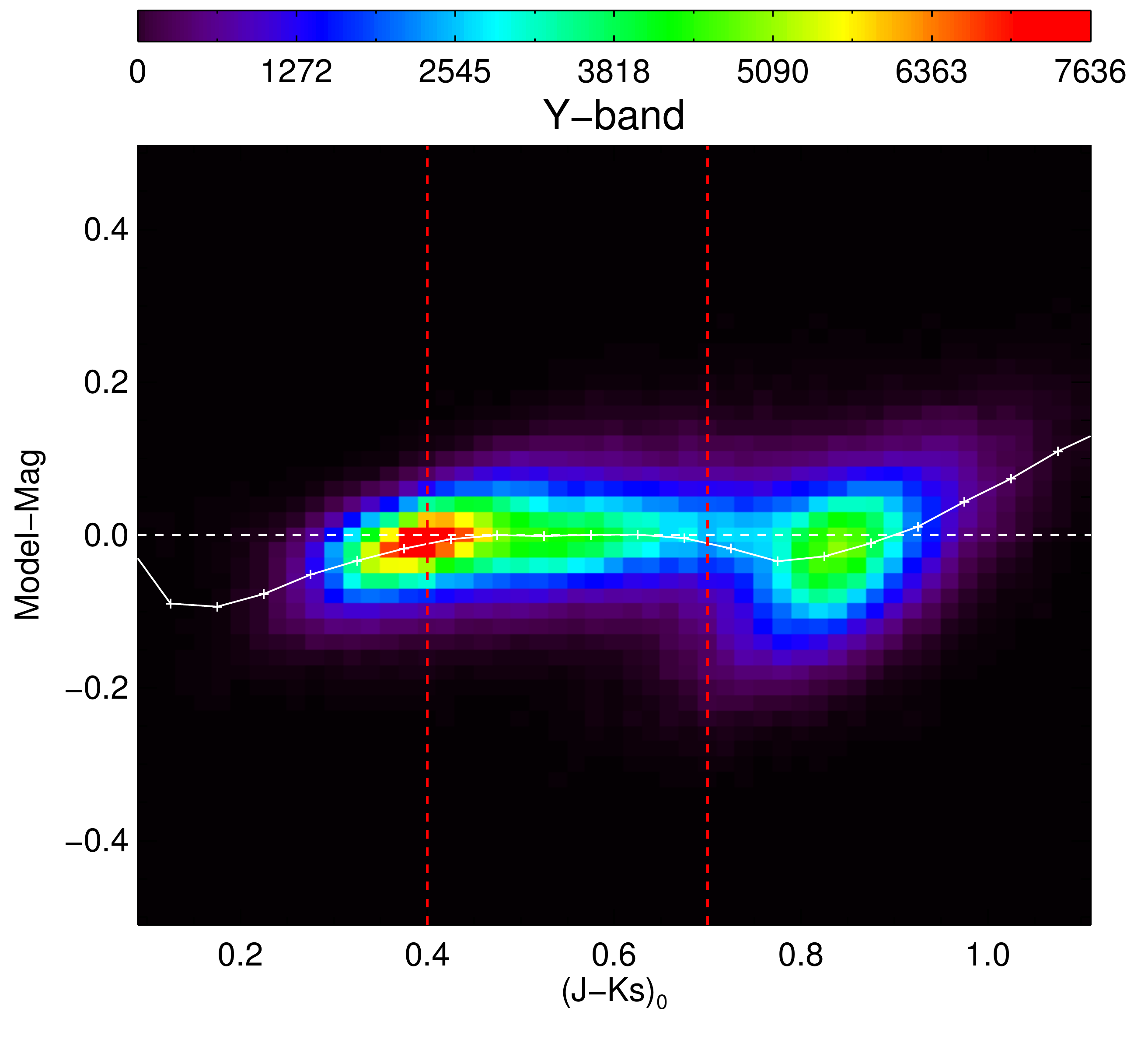} \\
\includegraphics[width=0.5\hsize,angle=0]{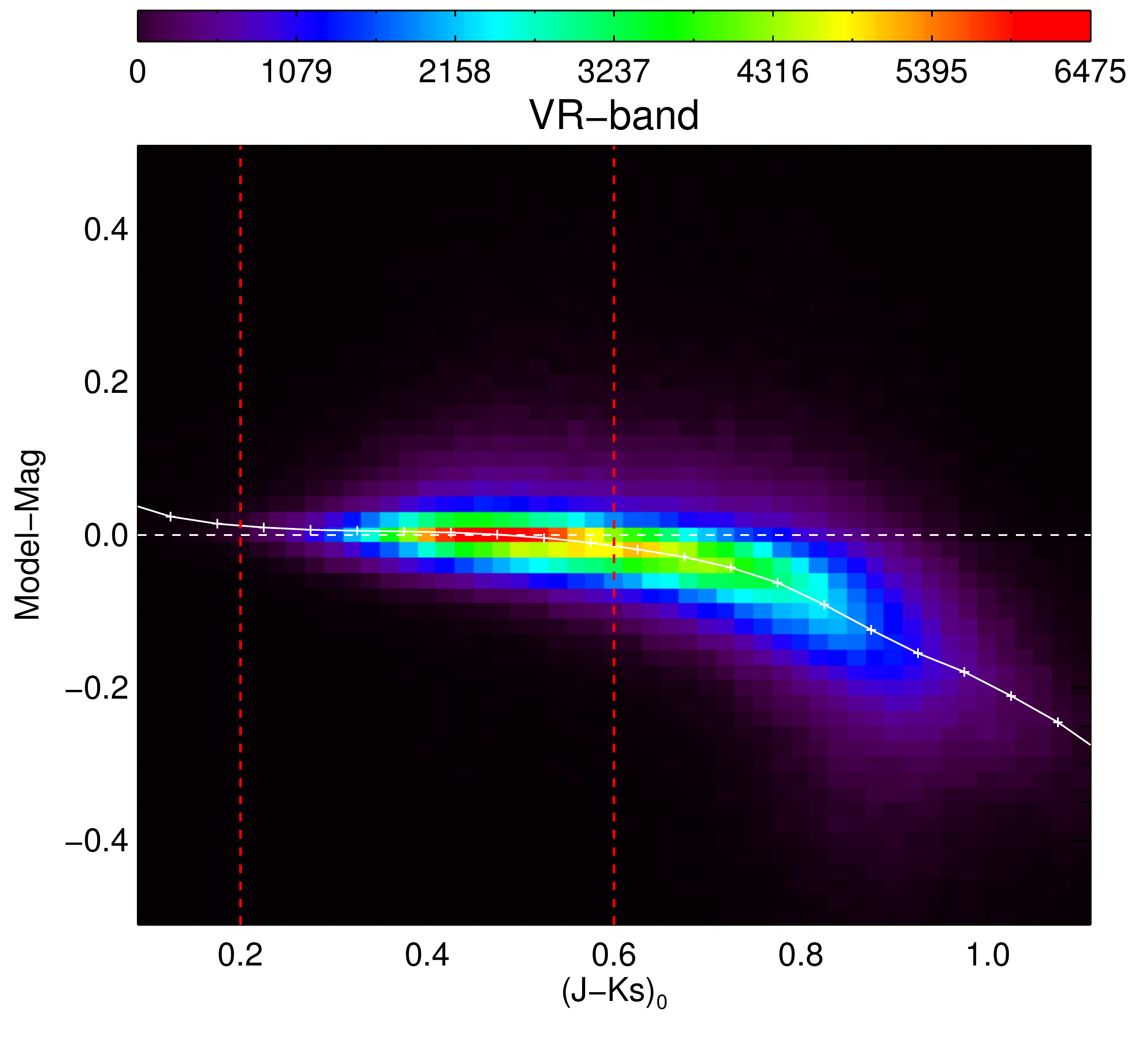}
\end{array}$
\caption{(cont.)}
\end{figure*}
\end{center}

\begin{enumerate}
\item The best model magnitudes were determined by comparing a linear combination of the photometry from the chosen public all-sky catalogs to a photometric reference catalog that set the zero point of the NSC photometric system (PS1 for $grizY$, SMASH for $u$-band and Gaia $G$ for {\em VR}). The coefficients were determined by comparing photometry in the Stripe82 region ($-$60\degr$<$$\alpha$$<$+60\degr, $-$1.1\degr$<$$\delta$$<$1.1\degr) for $grizY$ and {\em VR}, and in the region of the 41 fields of the SMASH survey for $u$-band.       
\item  The NSC exposures in the Stripe82 region were calibrated (and combined) using zero points derived with the model magnitude coefficients determined in step 1.  The calibrated NSC magnitudes were then compared to the model magnitudes and inspected for residual color terms that could arise from the differences in the PS1 and DECam filters. The model magnitude coefficients were adjusted to remove the residual patterns but the mean values were kept the same to remain on the (approximate) photometric system of the reference catalogs.
\end{enumerate}

\noindent
The residuals (from step 2) versus color are shown in Figure \ref{fig_modelmag} and the final model magnitude coefficients are presented in Table \ref{table_calib}.  Figure \ref{fig_modelmag} also shows the model magnitude color ranges used to determine the zero points which were determined by selecting the color region in the residuals where the systematics were small (excluding regions where the relations are non-linear or $\Delta{\rm m}\gtrsim$0.03 mag).

\begin{center}
\begin{figure}[t]
\includegraphics[width=1.0\hsize,angle=0]{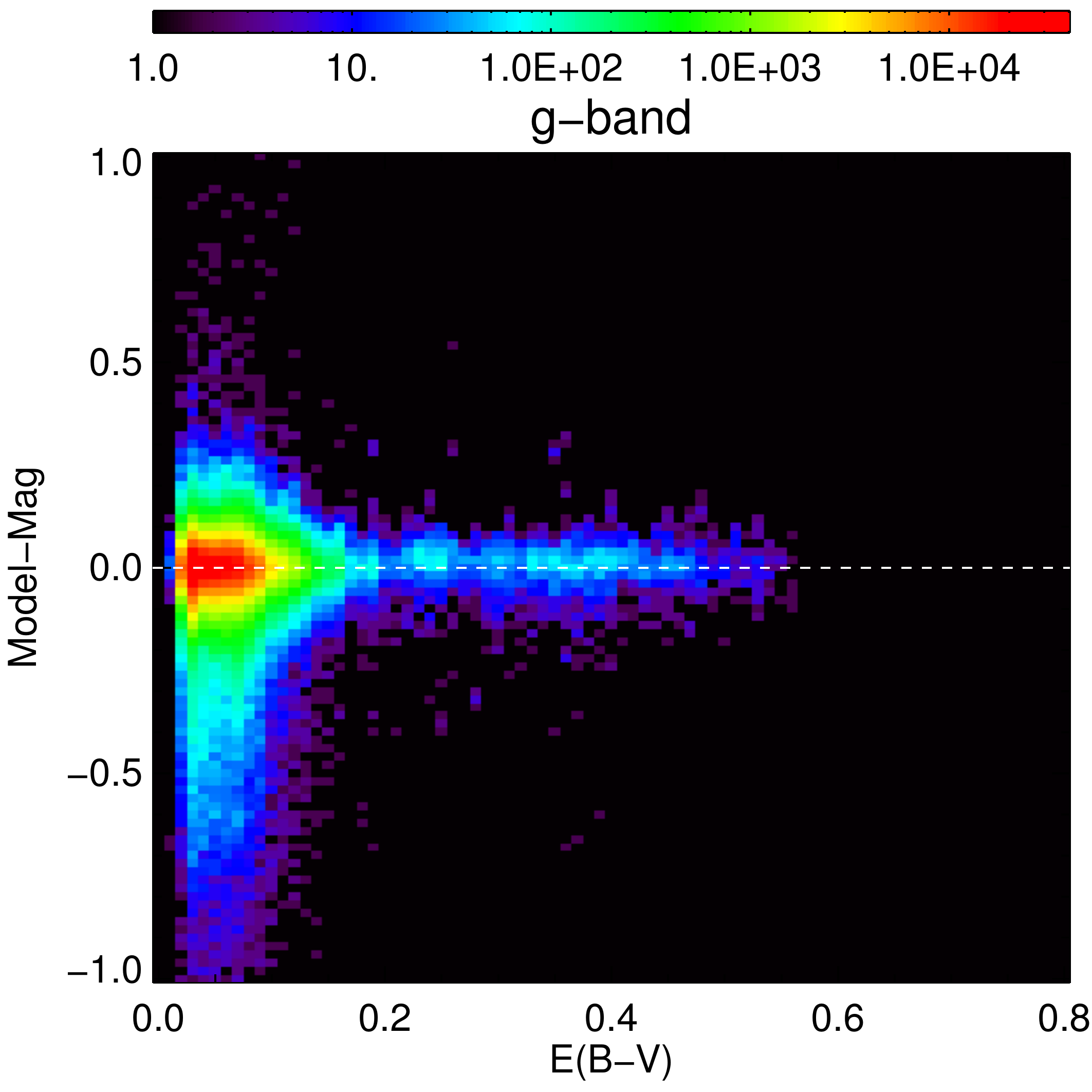}
\caption{Similar to Fig.\ \ref{fig_modelmag} but now showing an example of residual magnitude versus the SFD $E(B-V)$ reddening (for $g$-band).  The model magnitudes work well even in moderate extinction.}
\label{fig_modelmag_ebv}
\end{figure}
\end{center}

The photometric zero point of an image is estimated from the median of all calibration stars measured across all CCDs in an exposure. These 
``exposure-level'' zero points were used because the density of some of the catalogs needed to construct the model magnitudes was too low (especially APASS) to determine accurate chip-level zero points.  However, chip-level zero points were computed when possible ($\geq$5 reference objects per chip) in order to estimate spatial variations in zero points across an exposure that could indicate non-photometric conditions.  Even with exposure-level zero points, some exposures failed due to a lack of good photometric reference stars.  In these instances, well-calibrated, overlapping NSC exposures in the same filter were used to obtain the zero point ({\tt ZPTYPE}=2 in the {\tt exposure} table). Table \ref{table_zpstats} gives statistics on the zero point scatter and uncertainties per filter; the random uncertainties on the zero points are quite small.

Figure \ref{fig_modelmag_ebv} shows the residuals of the model magnitudes for $g$-band versus the SFD $E(B-V)$ reddening indicating that the model magnitudes work well in regions of moderate extinction.  The model magnitudes likely break down in very high extinction regions at the Milky Way mid-plane, as do the SFD reddening values. Hence, the photometric zero points might not be as reliable in these high-extinction regions.

\begin{center}
\begin{figure}[t]
\includegraphics[width=1.0\hsize,angle=0]{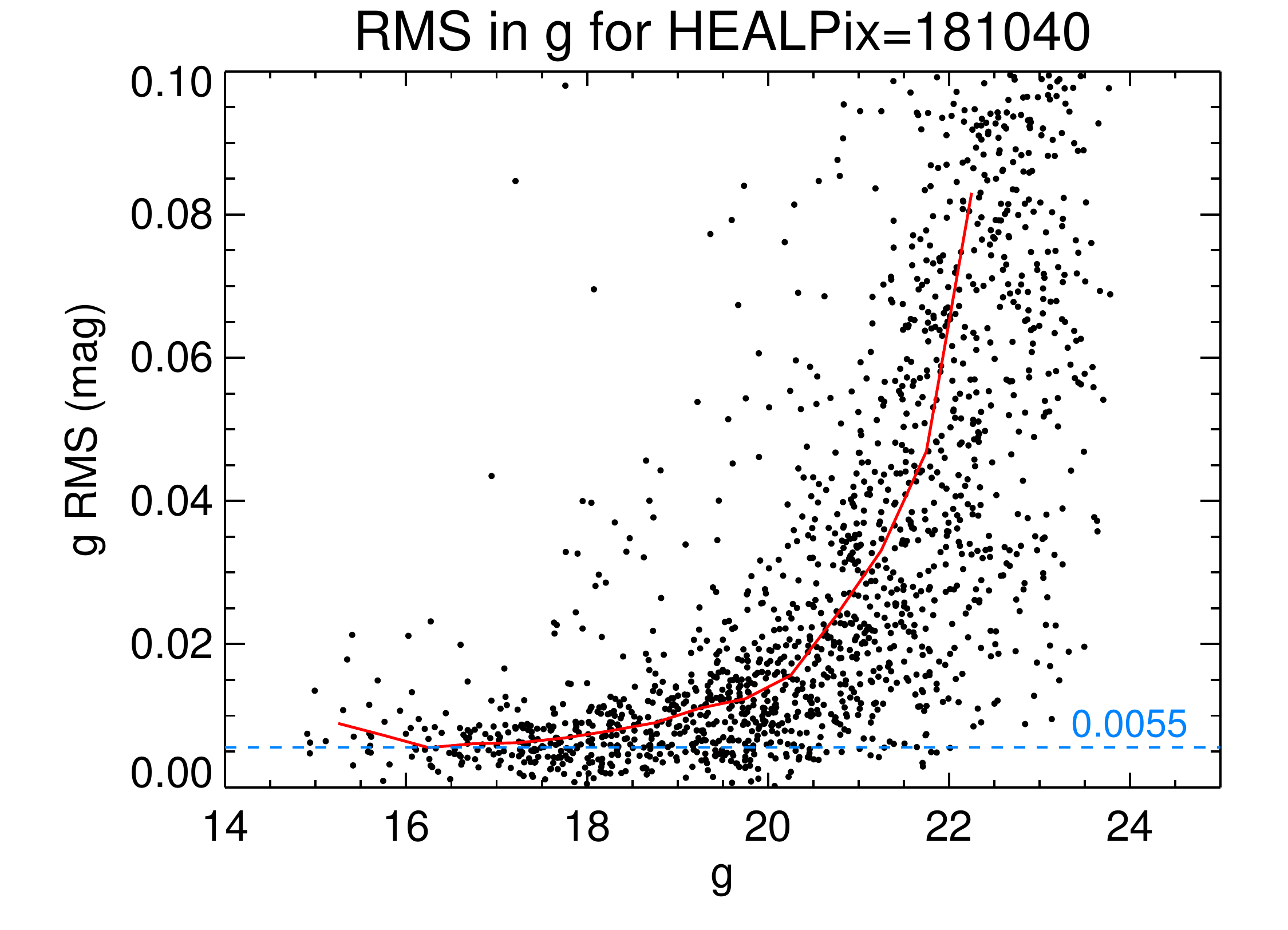}
\caption{The photometric RMS (over multiple detections) in $g$-band vs.~average $g$ magnitude for stars in the HEALPix=181040.  Median values in bins of 0.5 mag are shown in the red curve.  The lower plateau at 0.0055 mag is indicated by the blue dashed line.}
\label{fig_photscatter_mag}
\end{figure}
\end{center}

\begin{center}
\begin{figure}[t]
\includegraphics[width=1.0\hsize,angle=0]{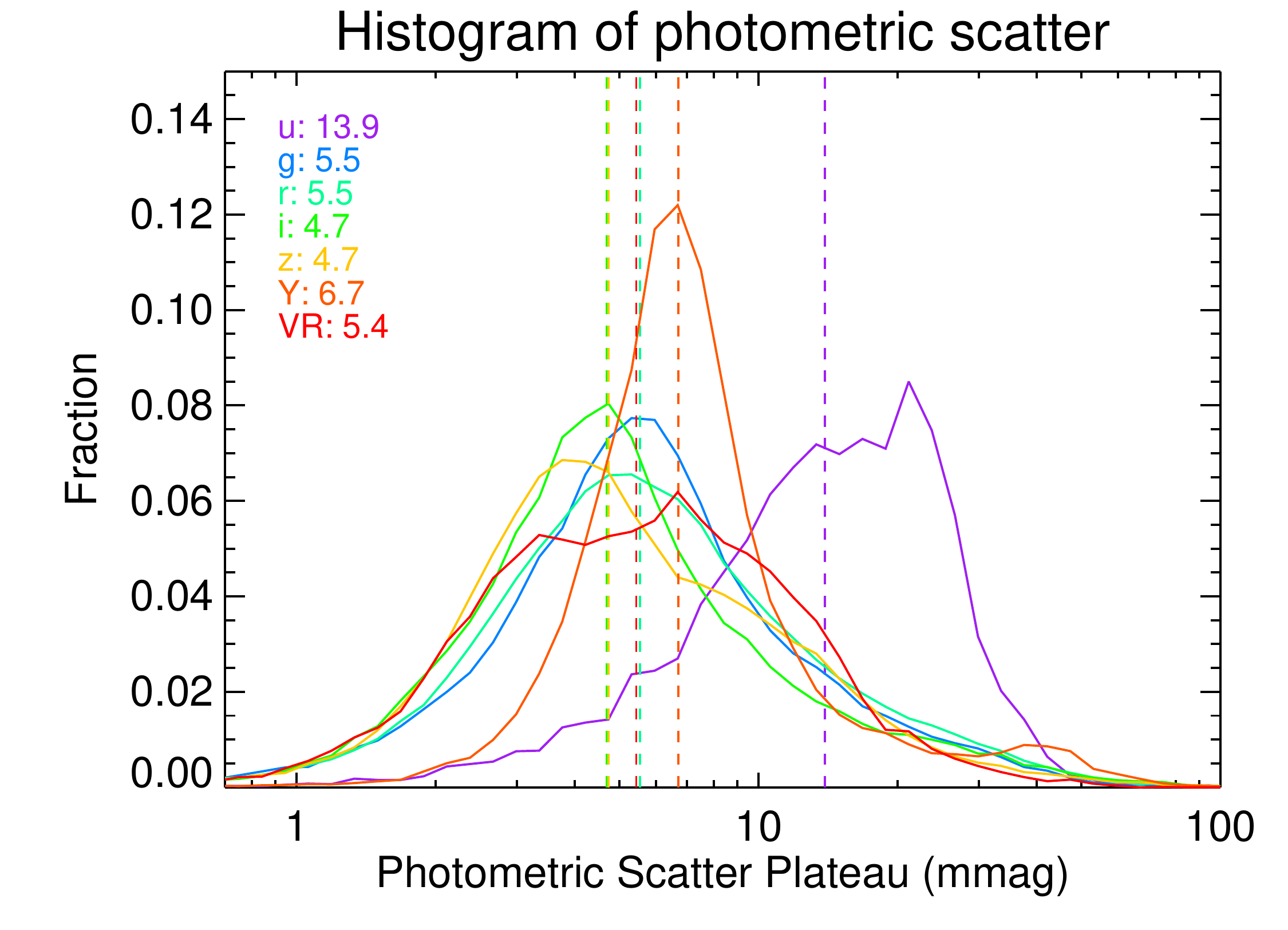}
\caption{The histogram of the plateau in the photometric RMS scatter of bright stars in each HEALPix for the seven bands. Median values are given in the legend and indicated by the vertical dashed lines.}
\label{fig_photscatter_hist}
\end{figure}
\end{center}

\begin{center}
\begin{figure*}[ht]
$\begin{array}{cc}
\includegraphics[trim={0cm 4.9cm 2cm 1cm},clip,width=0.50\hsize,angle=0]{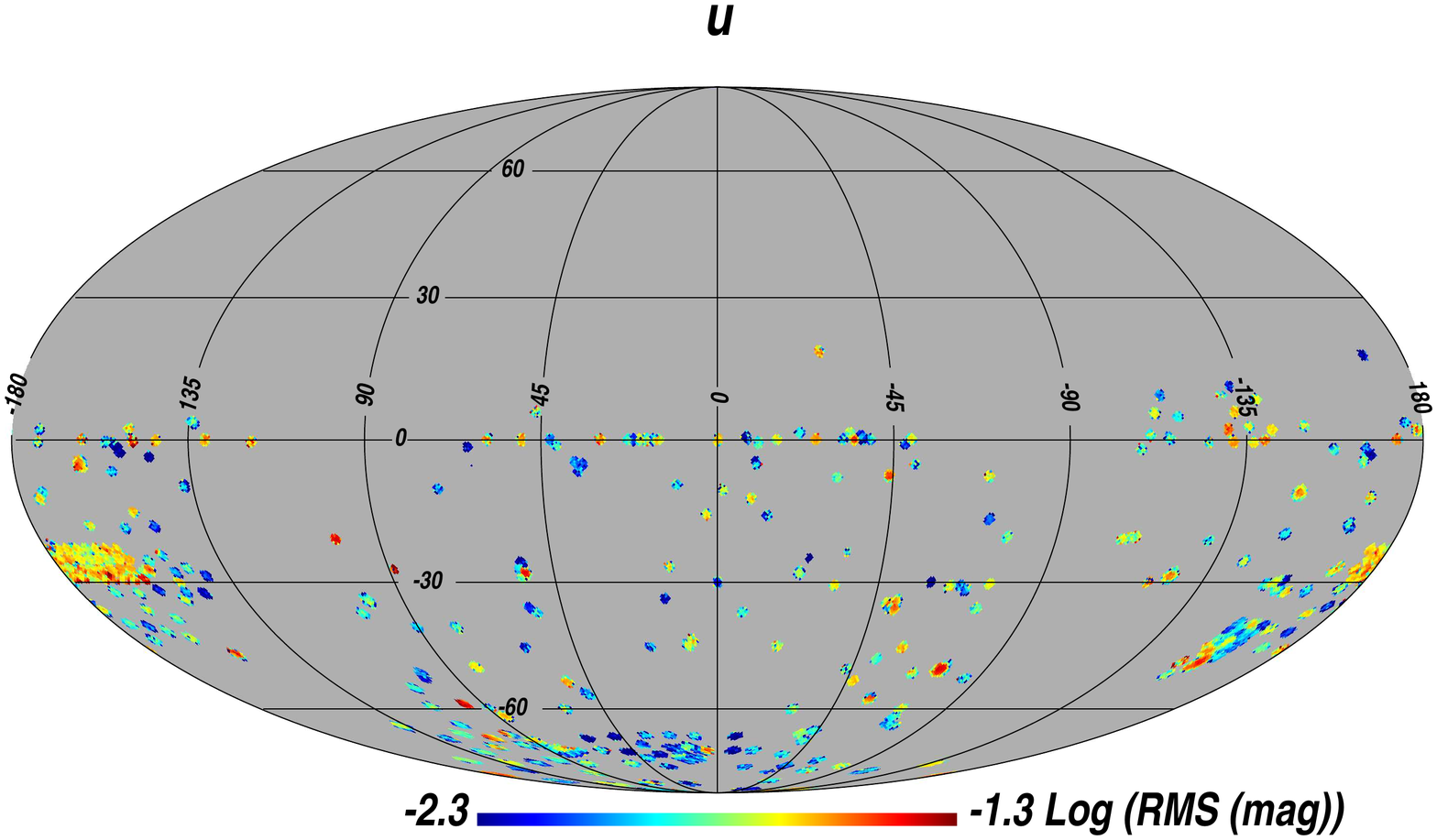}
\includegraphics[trim={0cm 4.9cm 2cm 1cm},clip,width=0.50\hsize,angle=0]{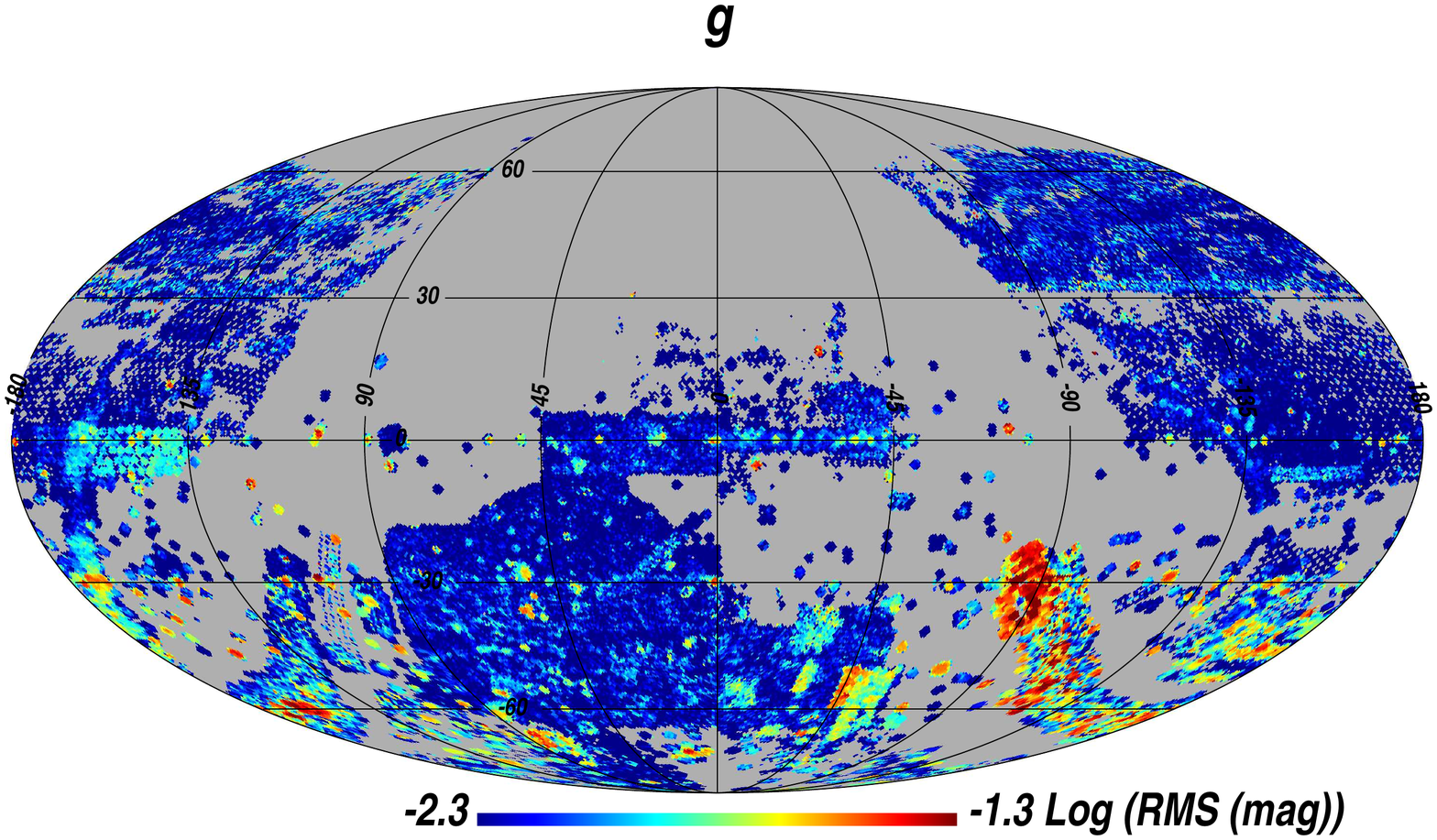} \\
\includegraphics[trim={0cm 4.9cm 2cm 1cm},clip,width=0.50\hsize,angle=0]{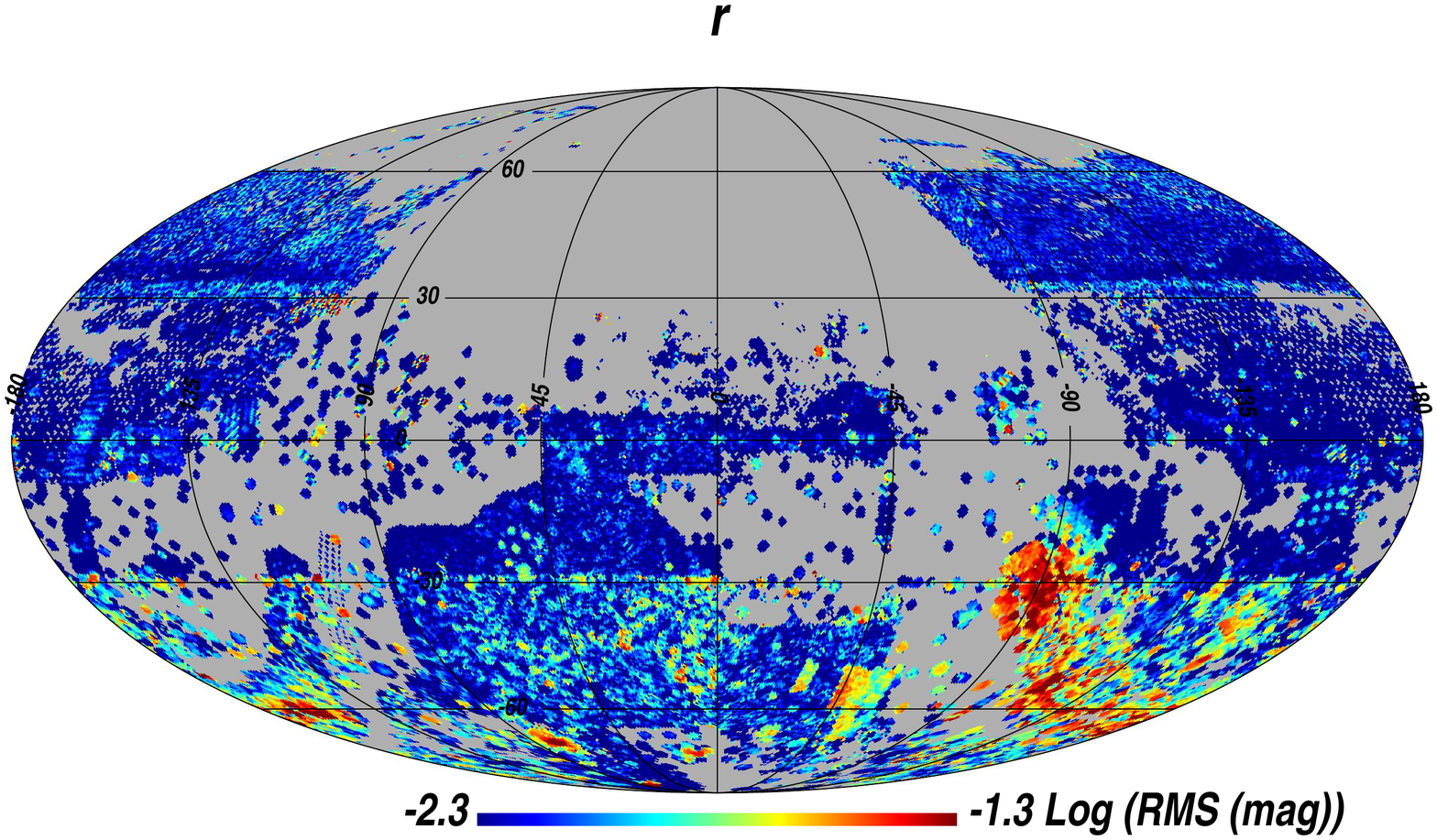}
\includegraphics[trim={0cm 4.9cm 2cm 1cm},clip,width=0.50\hsize,angle=0]{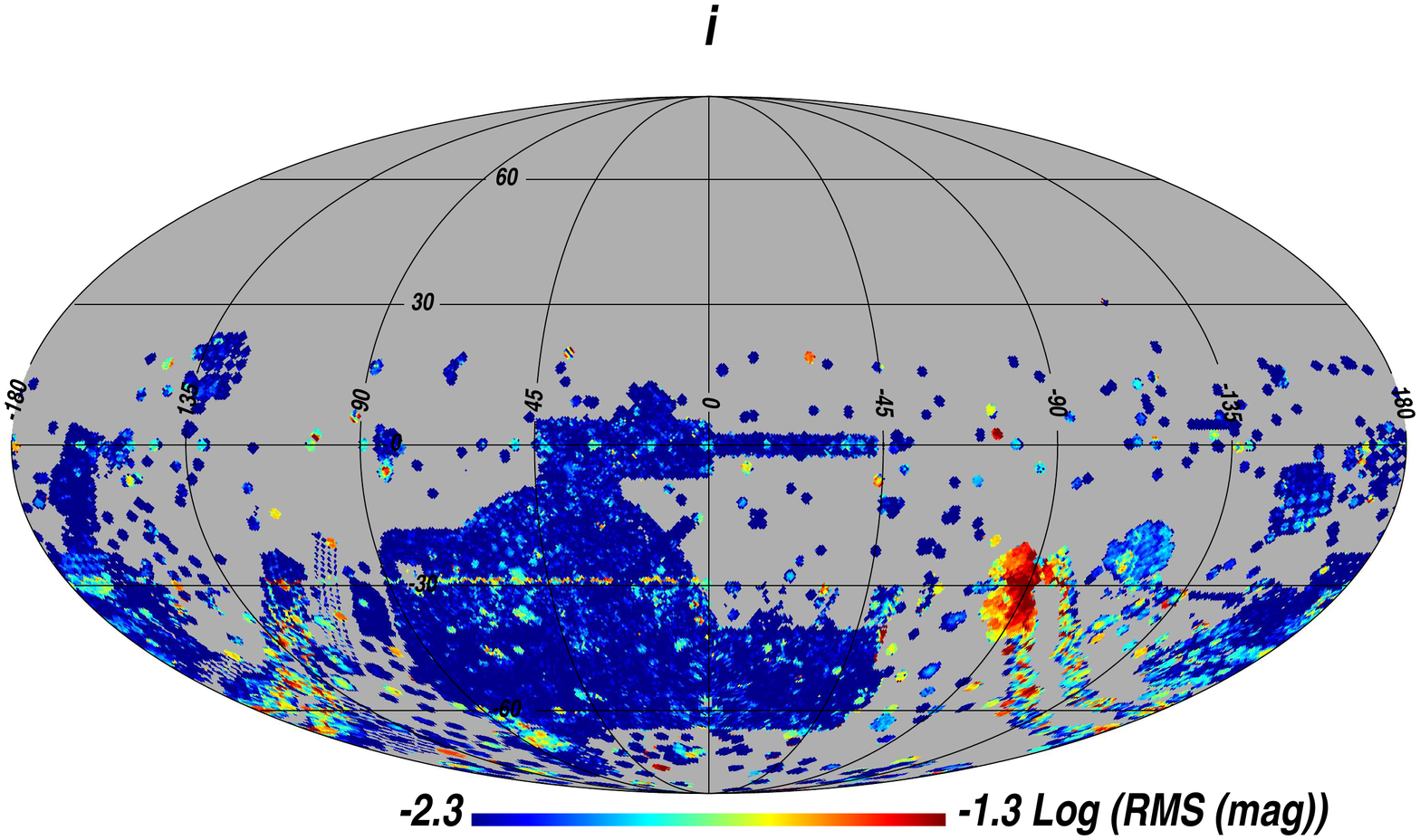} \\
\includegraphics[trim={0cm 4.9cm 2cm 1cm},clip,width=0.50\hsize,angle=0]{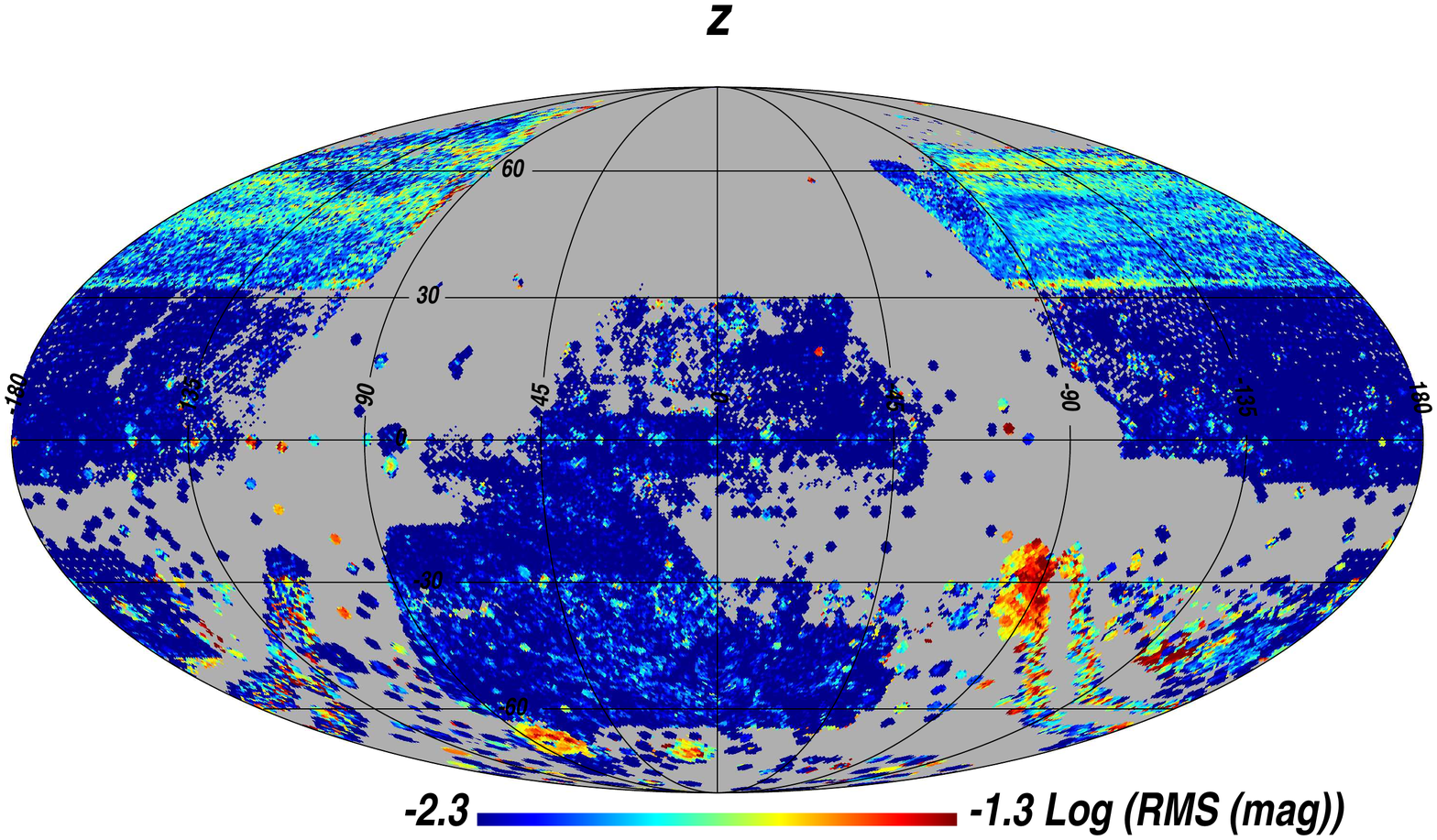}
\includegraphics[trim={0cm 4.9cm 2cm 1cm},clip,width=0.50\hsize,angle=0]{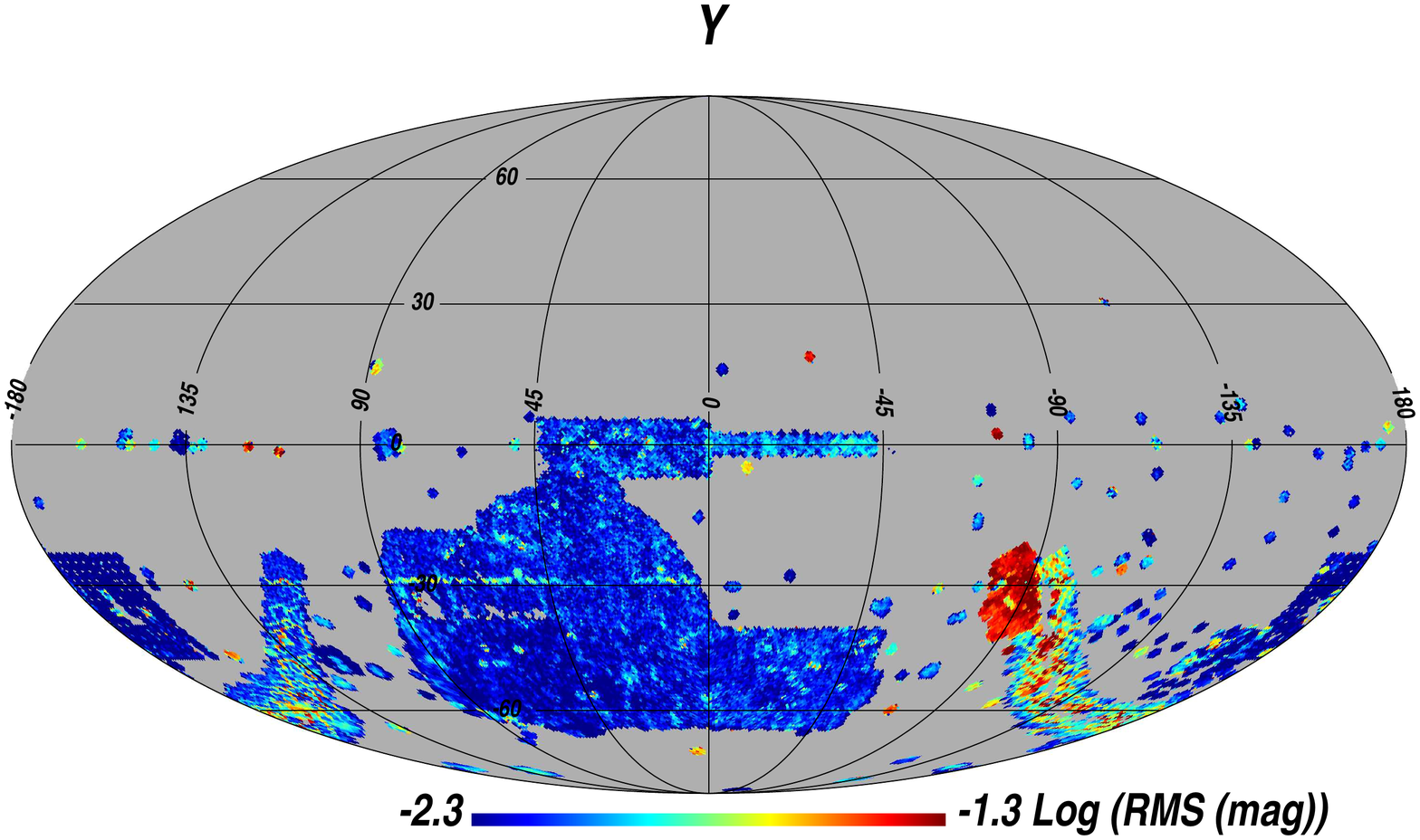} \\
\includegraphics[trim={0cm 4.9cm 2cm 1cm},clip,width=0.50\hsize,angle=0]{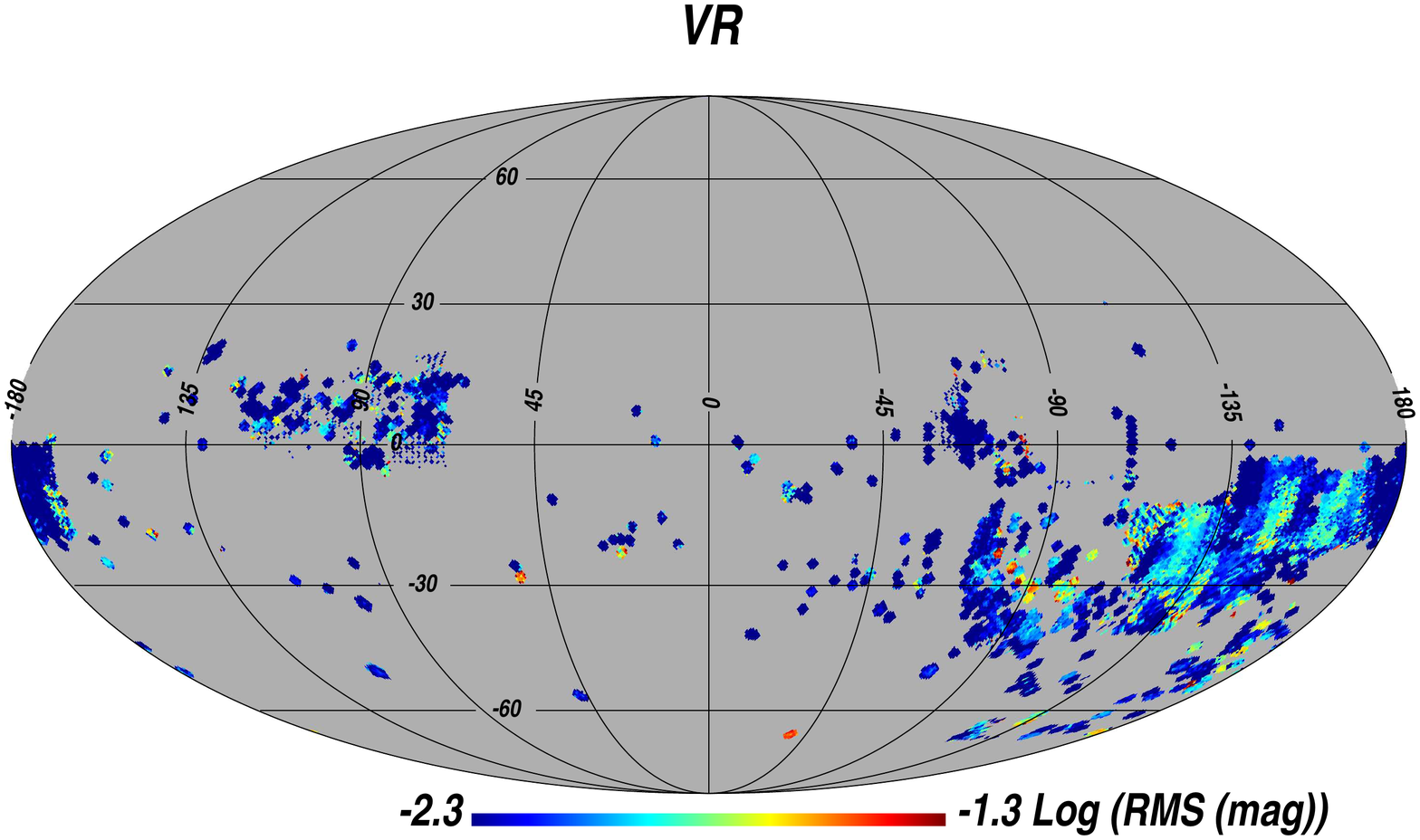}
\end{array}$
\caption{Maps of the photometric RMS scatter of bright stars (with more than two measurements) for the seven $u, g, r, i, z, Y$ and {\em VR} bands on a logarithmic scale in equatorial Aitoff projection.}
\label{fig_photscatter_maps}
\end{figure*}
\end{center}

The RMS in the photometry of bright stars from multiple detections is $\sim$1--2\%.  Figure \ref{fig_photscatter_mag} shows the photometric RMS for stars in one HEALPix region \citep{Gorski2005} with many exposures, illustrating that bright, unsaturated stars have low photometric scatter with a minimum in the binned medians of 0.0055 mag.
The histogram of RMS scatter values across all HEALPix pixels is shown in Figure \ref{fig_photscatter_hist}.  While there are variations, the distributions of all filters (besides $u$-band) are peaked below 0.01 mag indicating that the photometry is quite precise.
Figure \ref{fig_photscatter_maps} shows maps of the photometric RMS for the seven bands.  These maps show spatial variations in the RMS values.  In particular, the scatter is higher in the Galactic midplane (irrespective of whether PS1 or model magnitudes are used for determining the zero points) and the inner regions of the Magellanic Clouds, likely due to crowding effects on the photometric measurements.  The RMS is also slightly higher south of $\delta$=$-$29$^{\circ}$ (most noticable in $r$-band) likely due to the different calibration procedure using model magnitudes there.  The $z$-band scatter also appears to be somewhat higher for Mosaic-3 than DECam.

\begin{center}
\begin{figure*}[ht]
$\begin{array}{cc}
\includegraphics[trim={0cm 4.9cm 2cm 1cm},clip,width=0.50\hsize,angle=0]{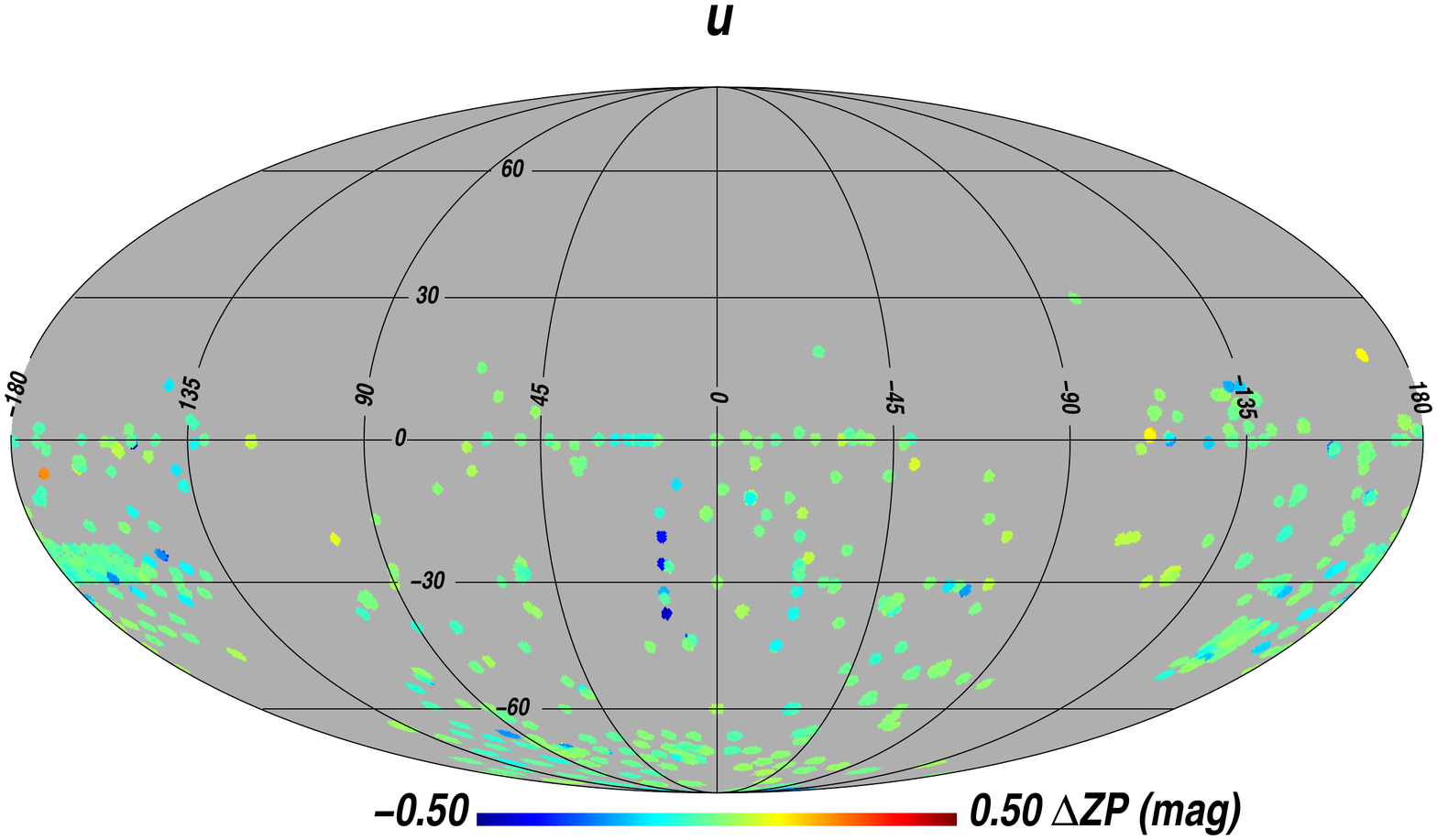}
\includegraphics[trim={0cm 4.9cm 2cm 1cm},clip,width=0.50\hsize,angle=0]{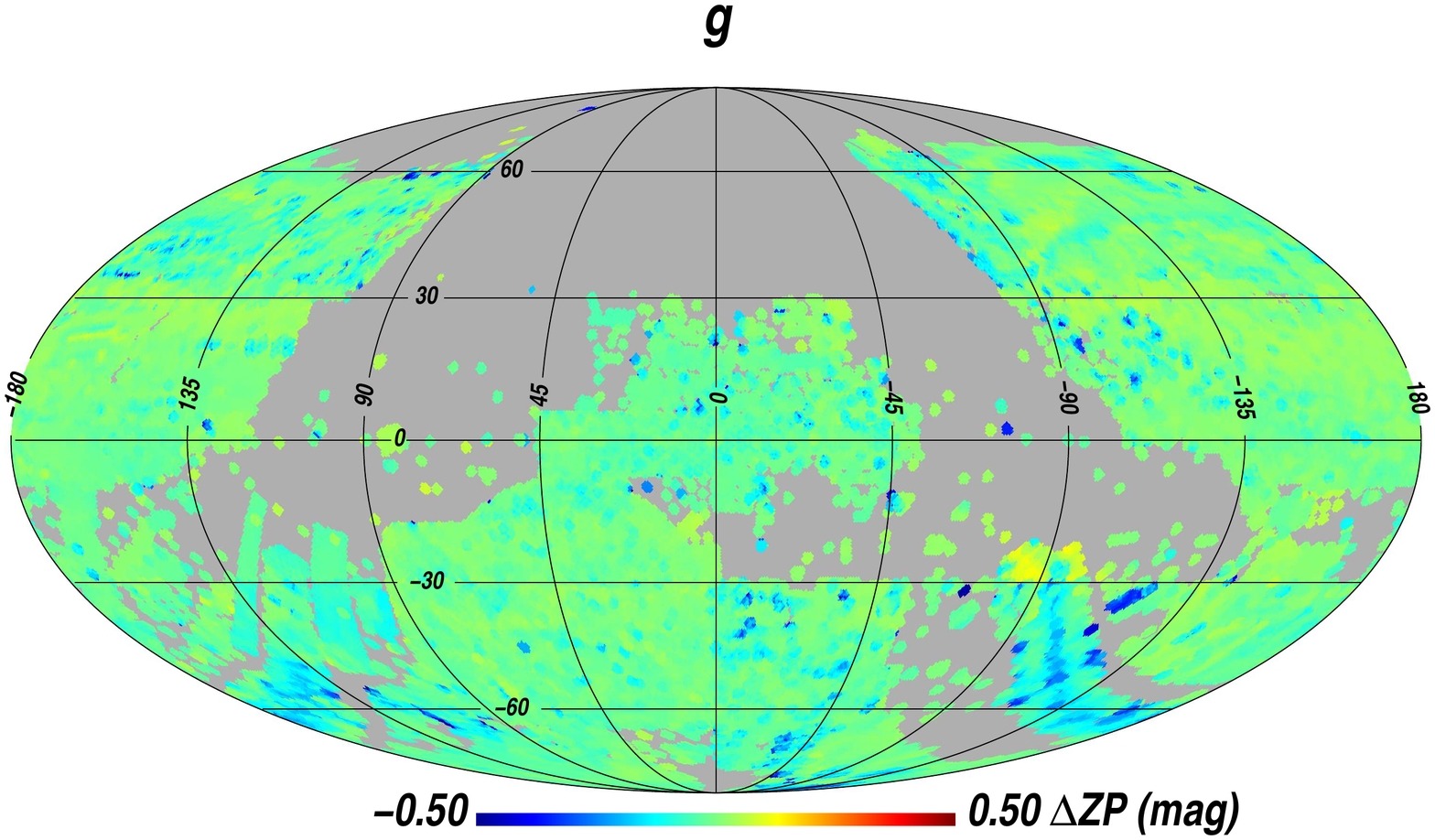} \\
\includegraphics[trim={0cm 4.9cm 2cm 1cm},clip,width=0.50\hsize,angle=0]{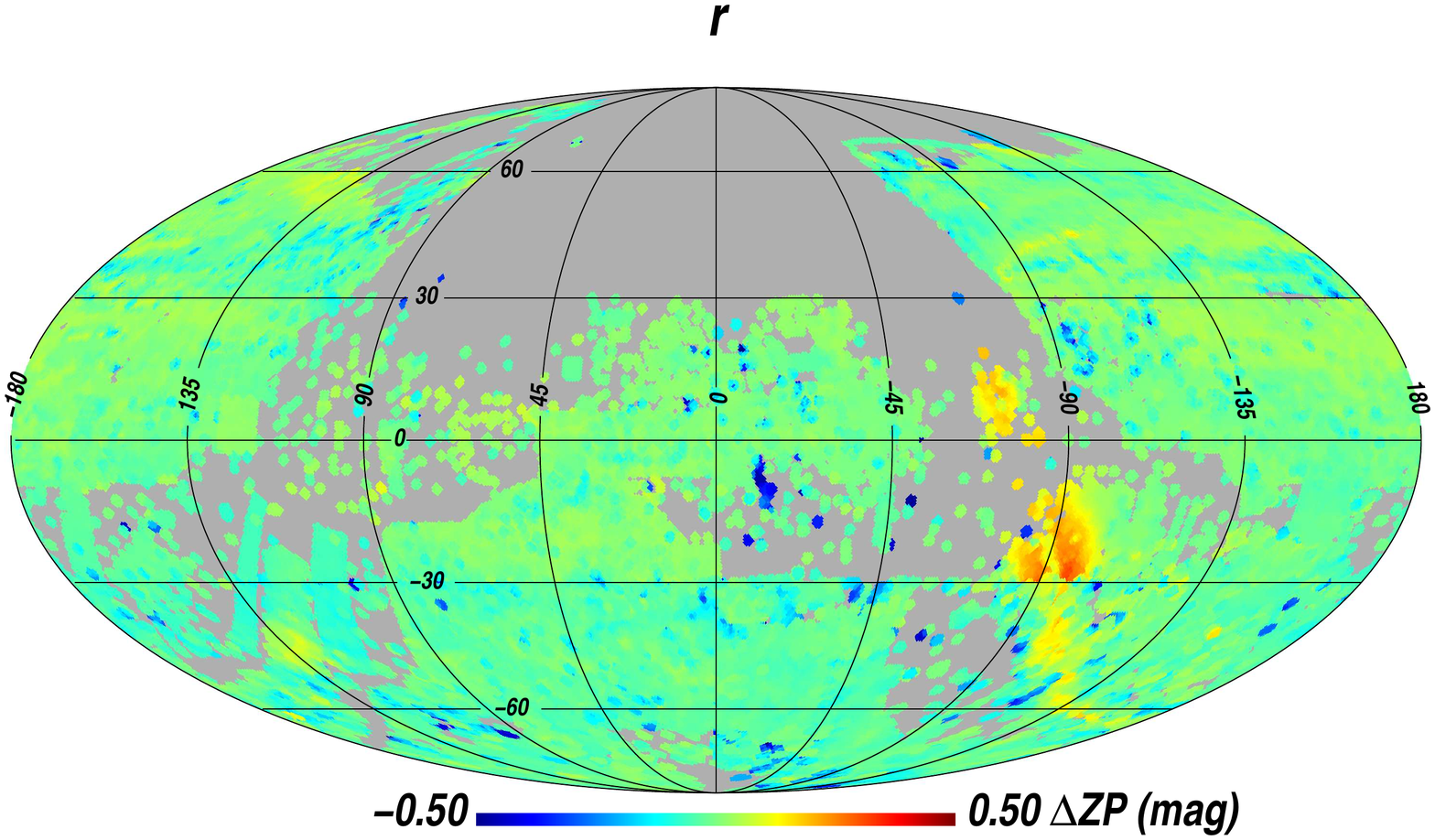}
\includegraphics[trim={0cm 4.9cm 2cm 1cm},clip,width=0.50\hsize,angle=0]{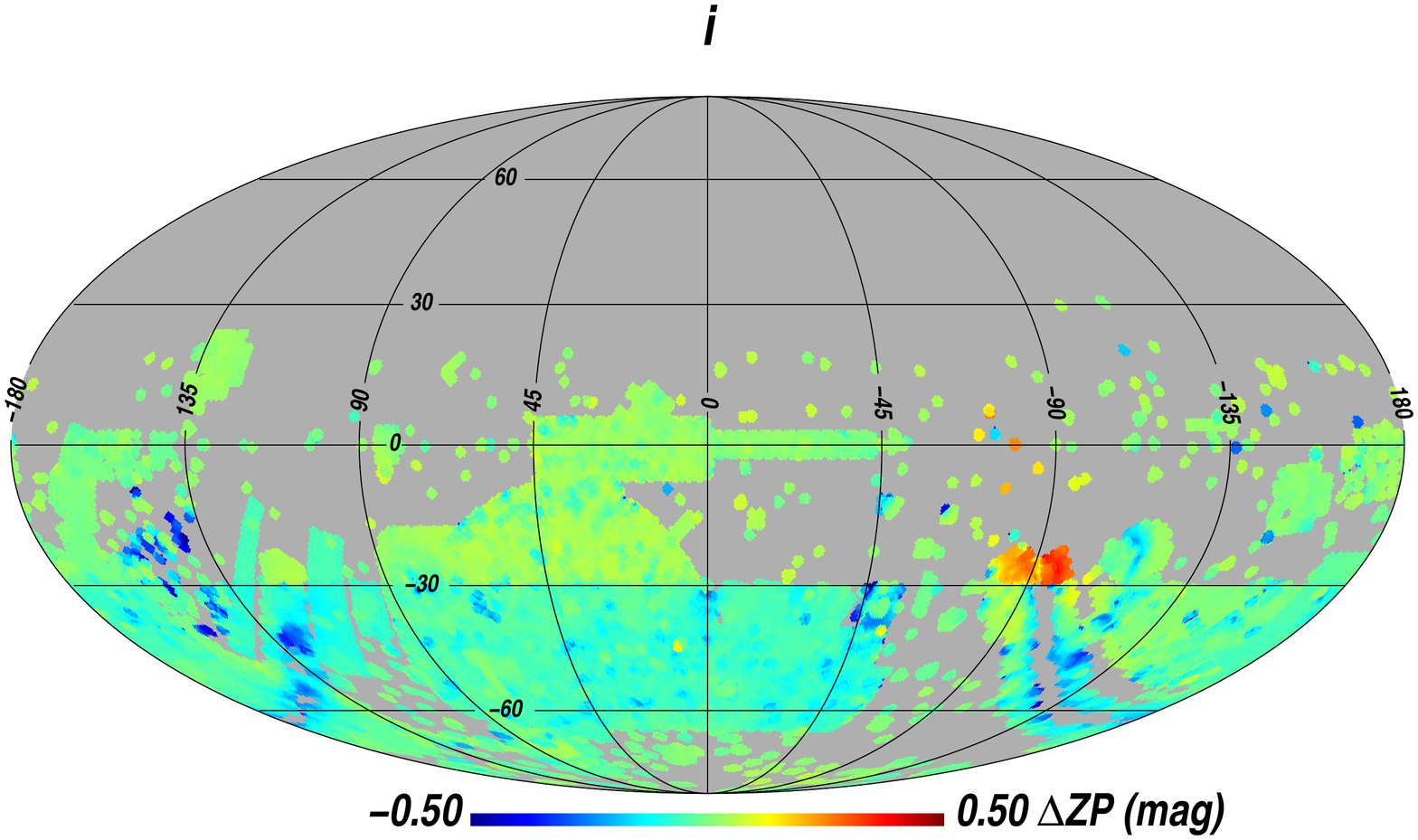} \\
\includegraphics[trim={0cm 4.9cm 2cm 1cm},clip,width=0.50\hsize,angle=0]{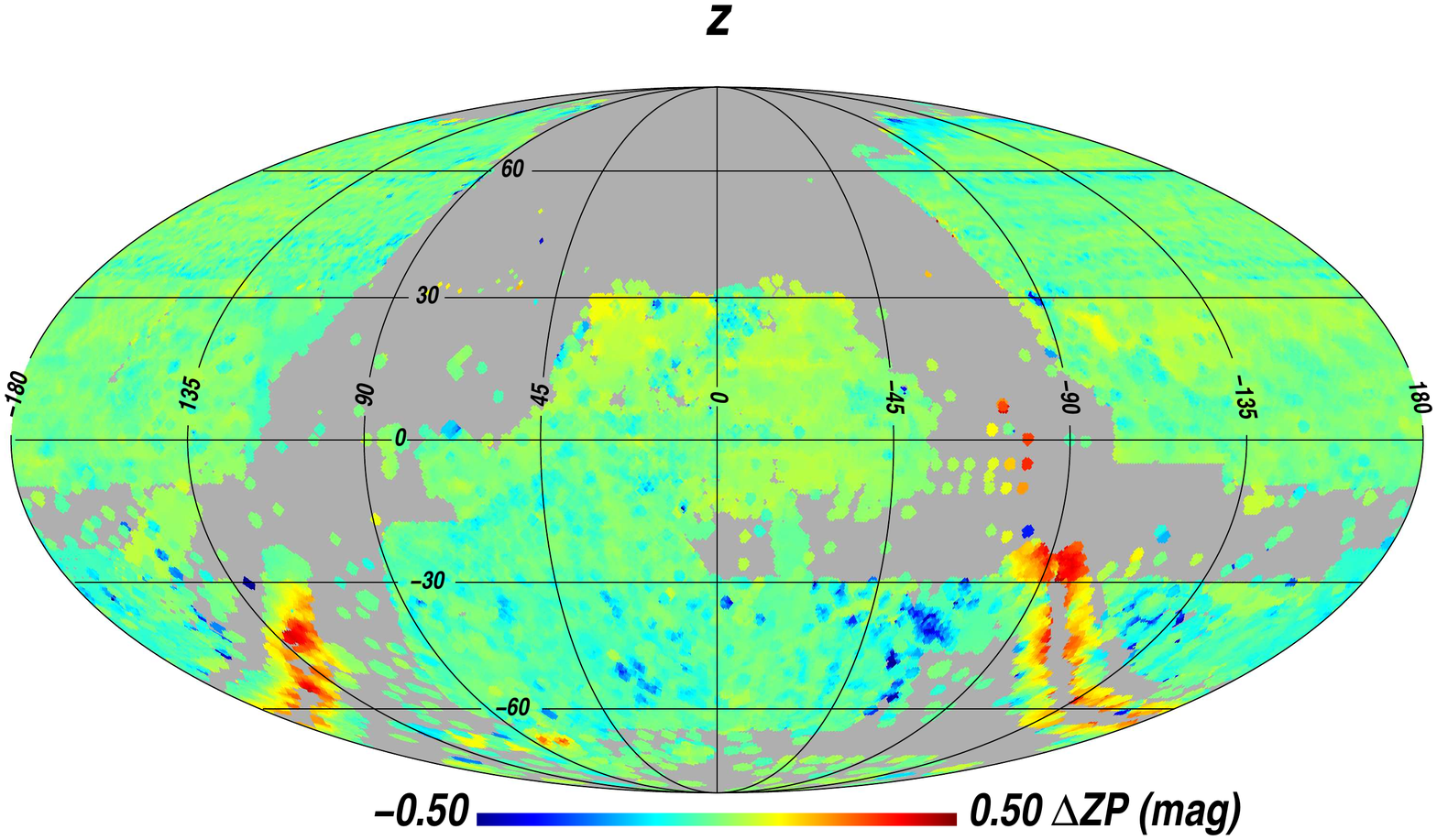}
\includegraphics[trim={0cm 4.9cm 2cm 1cm},clip,width=0.50\hsize,angle=0]{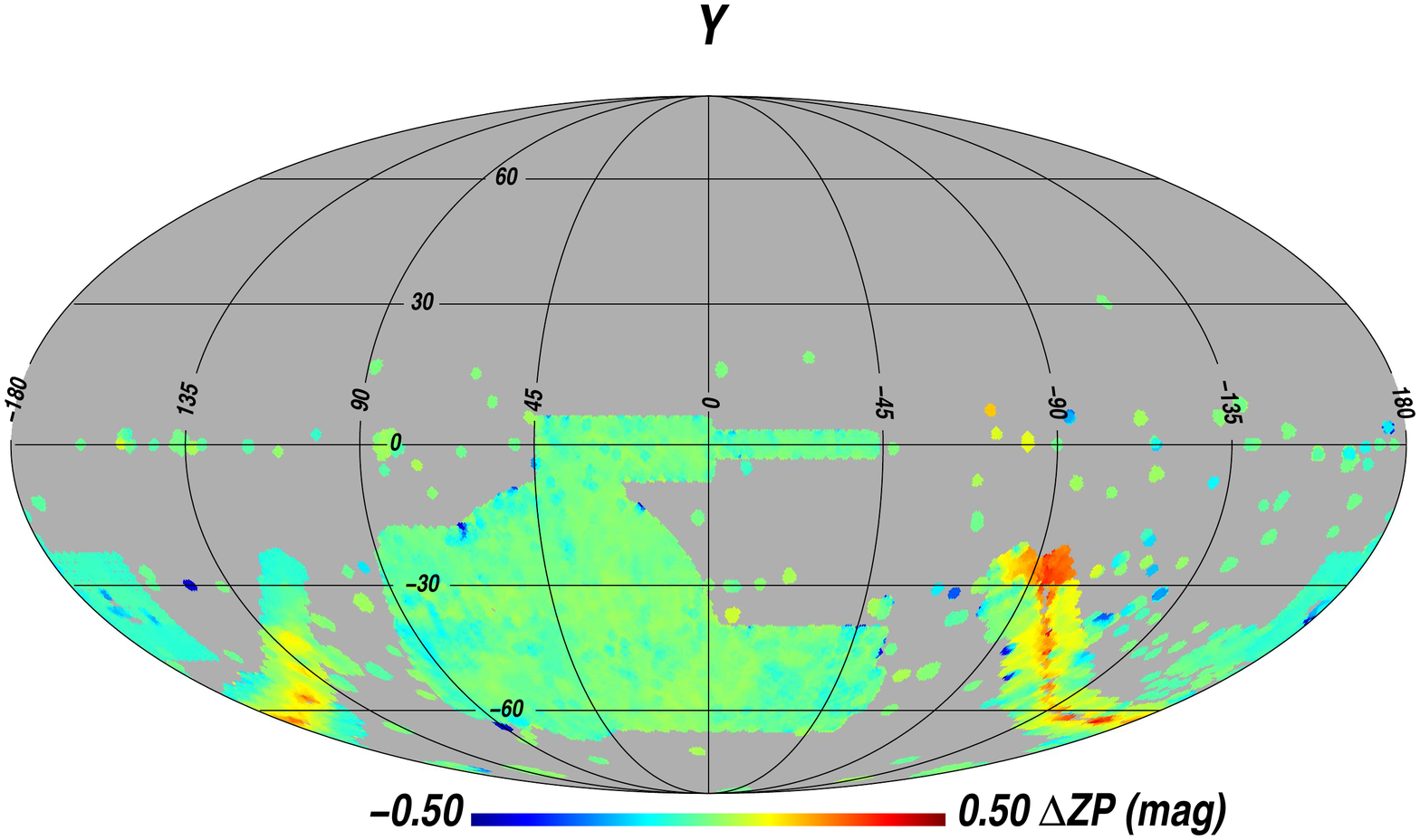} \\
\includegraphics[trim={0cm 4.9cm 2cm 1cm},clip,width=0.50\hsize,angle=0]{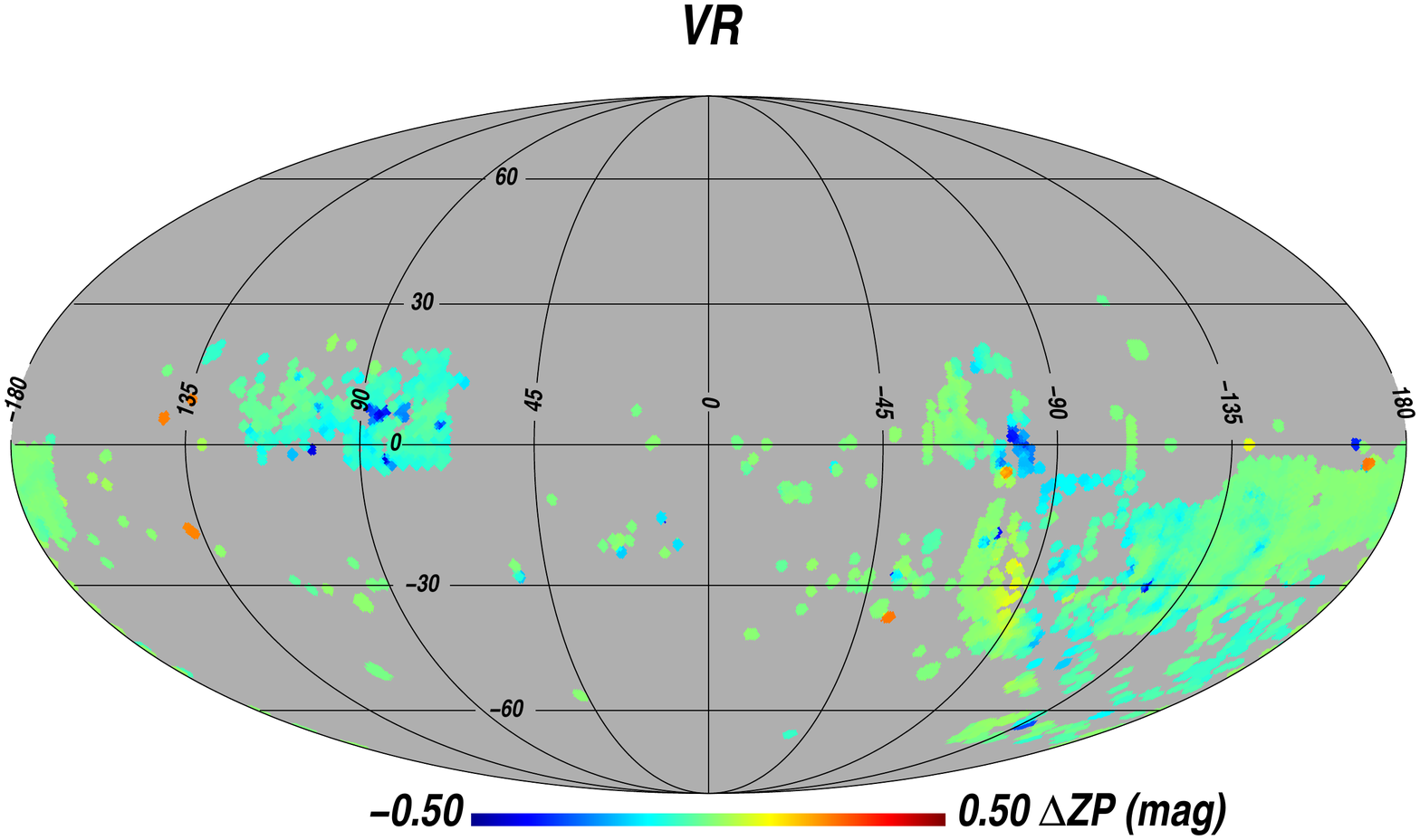}
\end{array}$
\caption{Maps of the mean photometric zero point in each Healpix relative to the mean across all exposures in a given band (in equatorial Aitoff projection).  The airmass-dependent extinction effects per exposure and long-term temporal variations in the zero points have been removed.}
\label{fig_zeropoint_maps}
\end{figure*}
\end{center}

Although the photometric RMS is small in most regions, this does not imply that the zero point is stable at that level across the entire sky.  There are known systematics at the $\approx$1\% level in the reference catalogs used to construct the model magnitudes \citep[e.g., 2MASS;][]{Magnier2013}.  In Figure \ref{fig_zeropoint_maps}, we show maps of our mean zero points across the sky, which reveal some spatially correlated systematics.
%is an attempt to investigate systematics in the photometric zero point across the sky. 
%It shows maps of the mean zero point in each HEALPix relative to the mean across all exposures in that band.
Airmass-dependent extinction effects and long-term temporal variations in the zero points have been removed but short-term variations have not.  The RMS across all HEALPix is $\approx$0.035 mag in most bands with $i$ and $z$-band being slightly higher (0.056/0.041 mag respectively).
While variations are expected in the zero points due to changes in weather conditions, these maps give a general sense of the reliability of the photometric zero points --- effectively an upper limit to spatial variations --- and where there might be some shortcomings.  In particular, the deviations are higher in the Galactic midplane both for $\delta$$>$$-$29\degr~where PS1 is used and for $\delta$$<$$-$29\degr~where model magnitudes are used.  These deviations are likely due to crowding effects, 
aperture corrections (which we effectively absorb into the zero point), and extinction issues.  While these deviations do not necessarily imply that the photometry is incorrect, there is a ``jump" across the $\delta$=$-$29\degr~boundary in the Milky Way mid-plane (near $\alpha$=260\degr) of a few tenths of a magnitude that suggests one side is systematically offset from the other.
We advise caution when using the photometry in high-extinction and crowded regions because there might be some offsets in the photometric zero points.
Finally, there is also a small jump across the $\delta$=$-$30\degr~boundary in the $i$-band for a large range in $\alpha$.  By comparing exposures near this boundary
calibrated with the two different methods as well as the mean zero point maps, we find that the model magnitudes ($\delta$$<$$-$29\degr) photometry in $i$-band is systematically brighter than the PS1 photometry by 0.065 mag.  It is not clear what the cause of this offset is, possibly an error when deriving the model
magnitude coefficients for this band.  None of the other bands show offsets at this level.

\begin{center}
\begin{deluxetable*}{lc}
\tablecaption{NSC Model Magnitude Equations}
\tablecolumns{2}
\tablehead{
  \colhead{Model Magnitude} & \colhead{Color Range}
}
\startdata
\vspace{0.1cm} 
$u$ = 0.247$\times$$NUV_{\rm GALEX}$ + 0.75$\times$$G_{\rm Gaia}$ + 0.54$\times$$(G-J)_0$ + 0.68$\times$$E(B-V)$ + 0.005    &    0.8$\le$$(G-J)_0$$\le$1.1 \\
\vspace{0.1cm} 
$g$ = $g_{\rm APASS}$ $-$ 0.0421$\times$$(J-K_{\rm s})_0$ $-$ 0.05$\times$$E(B-V)$ $-$ 0.0620  &    0.3$\le$$(J-K_{\rm s})_0$$\le$0.7 \\
\vspace{0.1cm} 
$r$ = $r_{\rm APASS}$ $-$ 0.086$\times$$(J-K_{\rm s})_0$ + 0.0$\times$$E(B-V)$ + 0.054    &       0.3$\le$$(J-K_{\rm s})_0$$\le$0.7 \\
\vspace{0.1cm} 
$i$ = $G_{\rm Gaia}$ $-$ 0.45$\times$$(J-K_{\rm s})_0$ - 0.27$\times$$E(B-V)$ + 0.096     &    0.25$\le$$(J-K_{\rm s})_0$$\le$0.65 \\
\vspace{0.1cm} 
$z$ = $J$ + 0.76$\times$$(J-K_{\rm s})_0$ + 0.40$\times$$E(B-V)$ + 0.60  &   0.4$\le$$(J-K_{\rm s})_0$$\le$0.65 \\
\vspace{0.1cm} 
$Y$ = $J$ + 0.54$\times$$(J-K_{\rm s})_0$ + 0.20$\times$$E(B-V)$ + 0.66   &       0.4$\le$$(J-K_{\rm s})_0$$\le$0.7 \\
\vspace{0.1cm} 
{\em VR} = $G_{\rm Gaia}$                            &      0.2$\le$$(J-K_{\rm s})_0$$\le$0.6 \\
\hline \\
\vspace{0.1cm} 
$(G-J)_0$ = $G_{\rm Gaia}$ $-$ $J$ $-$ 1.12$\times$$E(B-V)$ & \\
$(J-K_{\rm s})_0$ = $J$ $-$ $K_{\rm s}$ - 0.17$\times$$E(B-V)$ & 
\enddata
\label{table_calib}
\end{deluxetable*}
\end{center}

\begin{center}
\begin{deluxetable}{lcccc}
\tablecaption{Zero Point Statistics}
\tablecolumns{5}
\tablehead{
  \colhead{Filter} & \colhead{$\delta$ Range} & \colhead{Median} & \colhead{Median}
  & \colhead{Median} \\
    &  & \colhead{ZP RMS} & \colhead{ZP Error} & \colhead{N$_{\rm reference}$}
}
\startdata
$u$  &   all   &  0.064 &  0.0099   &     129 \\
$g$  & $>-$29  &  0.041 &  0.0007   &    3994 \\
$g$  & $<-$29  &  0.032 &  0.0043   &     439 \\
$r$  & $>-$29  &  0.057 &  0.0010  &    5829 \\
$r$  & $<-$29  &  0.049 &  0.0080   &     254 \\
$i$  & $>-$29  &  0.082 &  0.0013   &    5646 \\
$i$  & $<-$29  &  0.056 &  0.0017   &    1370 \\
$z$  & $>-$29  &  0.076 &  0.0027   &    1749 \\
$z$  & $<-$29  &  0.078 &  0.0027   &     898 \\
$Y$  & $>-$29  &  0.037 &  0.0007  &    4144 \\
$Y$  & $<-$29  &  0.055 &  0.0017   &     968 \\
{\em VR} &  all    &  0.017 &  0.0007  &     804
\enddata
\label{table_zpstats}
\end{deluxetable}
\end{center}

\subsection{Combination}
\label{subsec:combination}

The single-epoch measurement catalogs are crossmatched and combined in the final stage of the NSC processing.  A HEALPix scheme \citep{Gorski2005} is used (nside=128 with $\approx$0.21 square degrees per pixel and ring ordering) to fully tile the sky and allow for more efficient parallelization of the combination step.  Before the processing is initiated, exposure-level quality cuts are applied to all exposures that successfully completed the calibration step. We select:
\begin{itemize}
\item public data (as of 2017-10-11);
\item successfully astrometrically calibrated by the CP;
\item all chips astrometrically calibrated (using Gaia DR1) in NSC calibration step;
\item median $\alpha$/$\delta$ RMS across all chips $\leq$0.15\arcsec;
\item seeing FWHM $\leq$2\arcsec;
\item zero point (corrected for airmass extinction) within 0.5 mag of the temporally-smoothed\footnote{The zero points were B-spline smoothed over $\approx$200 nights to track system throughput variations.} zero point for that band;
\item zero point uncertainty $\leq$0.05 mag;
\item number of photometric reference stars $\geq$5;
\item spatial variation (RMS across chips) of zero point $\leq$0.1 mag
(only for DECam with number of chips with well-measured chip-level zero points $>$5);
\item not in a survey's bad exposure list (currently only for the Legacy Surveys and SMASH data).
\end{itemize}

\noindent
%Then the final list of exposures and the healpix pixels that they overlap is created.

For any given HEALPix pixel, all exposures overlapping it and its neighboring pixels are successively loaded.  Only sources of an exposure that fall within 10\arcsec~of the HEALPix pixel boundary are kept.  Single-epoch level quality cuts are also applied, and we only selected sources:
\begin{itemize}
\item with no CP mask flags set;
\item with no SExtractor object or aperture truncation flags;
\item not detected on the bad amplifier of DECam CCDNUM 31 (if MJD$>$56,600 or big background jump between amplifiers);
\item with S/N$\geq$5.
\end{itemize}

Sources are successively crossmatched (with a 0.5\arcsec~matching radius) to existing ``objects" or are added as new objects if no match is found.  Cumulative values of quantities are accumulated throughout this process and converted to average values at the very end. Most quantities such as morphology parameters are simple averages while coordinates and magnitudes per band are S/N-weighted averages.  Average morphology values are computed per band and across all bands.  Proper motions are computed as S/N-weighted least-squared fits but it should be noted that these are only relative to the foreground stars that are used to astrometrically calibrate each exposure to Gaia DR1.  Also, the proper motions are effectively limited to $\lesssim$200 mas yr$^{-1}$ (although this depends on the cadence) because the multiple measurements from higher proper motion stars are not matched to a common object owing to the 0.5\arcsec~matching radius. Photometric RMS values per band across the multiple measurements are also computed and can be used to ascertain the photometric variability of an object.  Only objects with final average coordinates within the boundary of the HEALPix pixel are kept.

%\section{Quality Assurance}
%\label{sec:qa}

% the check I did of photometric stability of zeropoints in a healpix with many exposures
% comparison with other catalogs, PS1
% astrometry check with Gaia, etc.

\section{Description and Achieved Performance of Final Catalog}
\label{sec:dr1}

\begin{center}
\begin{figure*}[ht]
$\begin{array}{cc}
\includegraphics[trim={0cm 5cm 2cm 1cm},clip,width=0.50\hsize,angle=0]{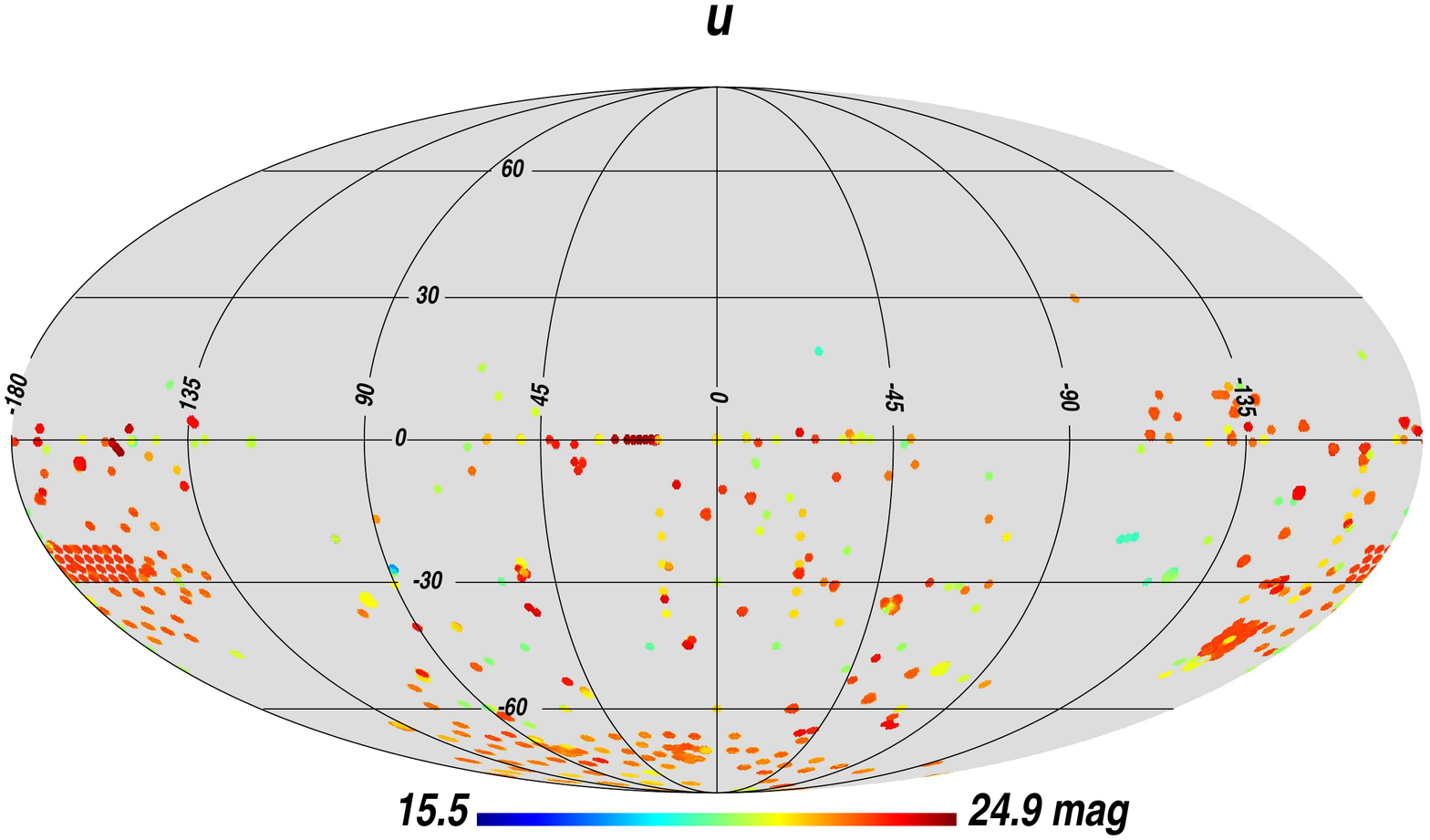}
\includegraphics[trim={0cm 5cm 2cm 1cm},clip,width=0.50\hsize,angle=0]{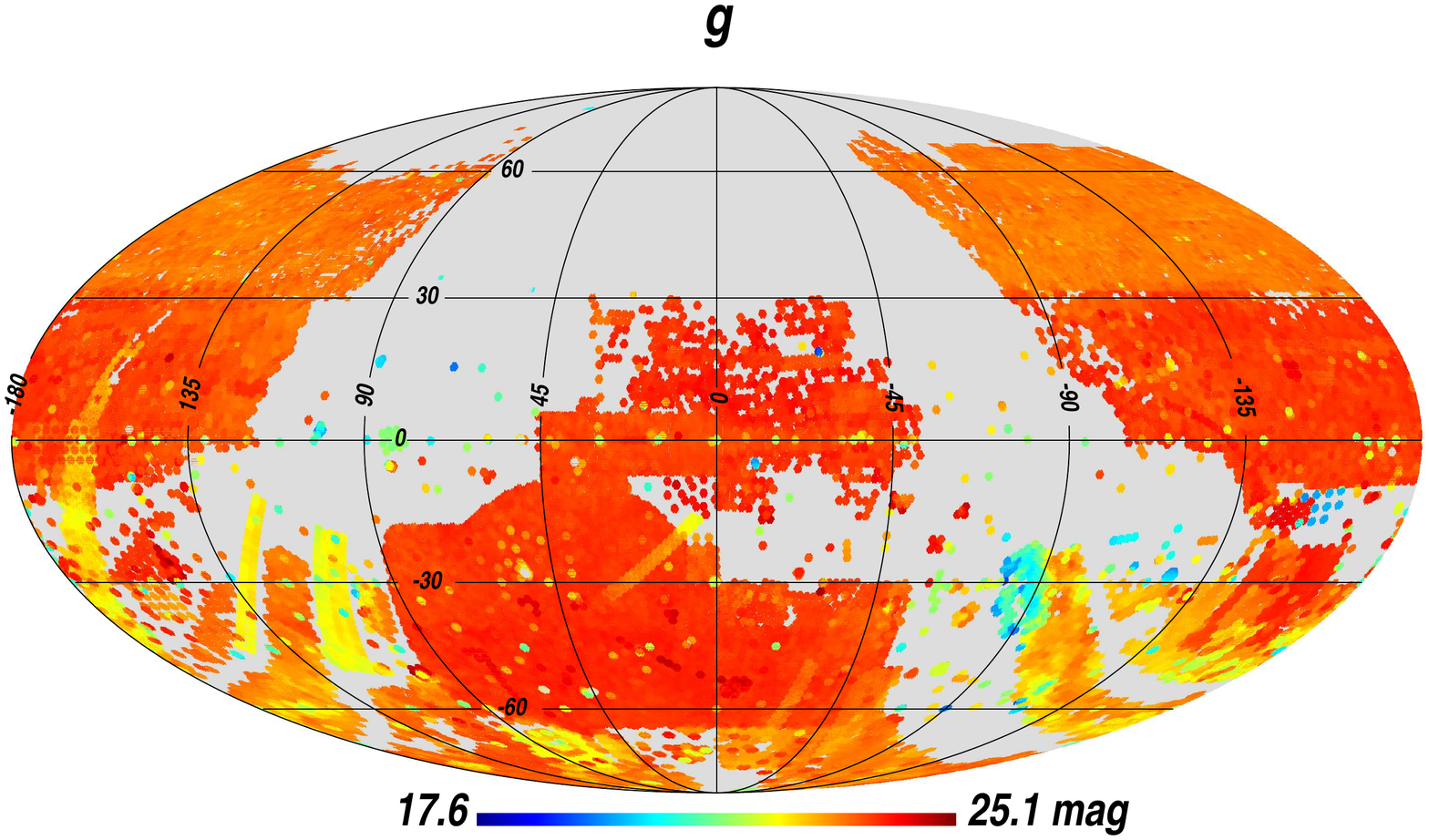} \\
\includegraphics[trim={0cm 5cm 2cm 1cm},clip,width=0.50\hsize,angle=0]{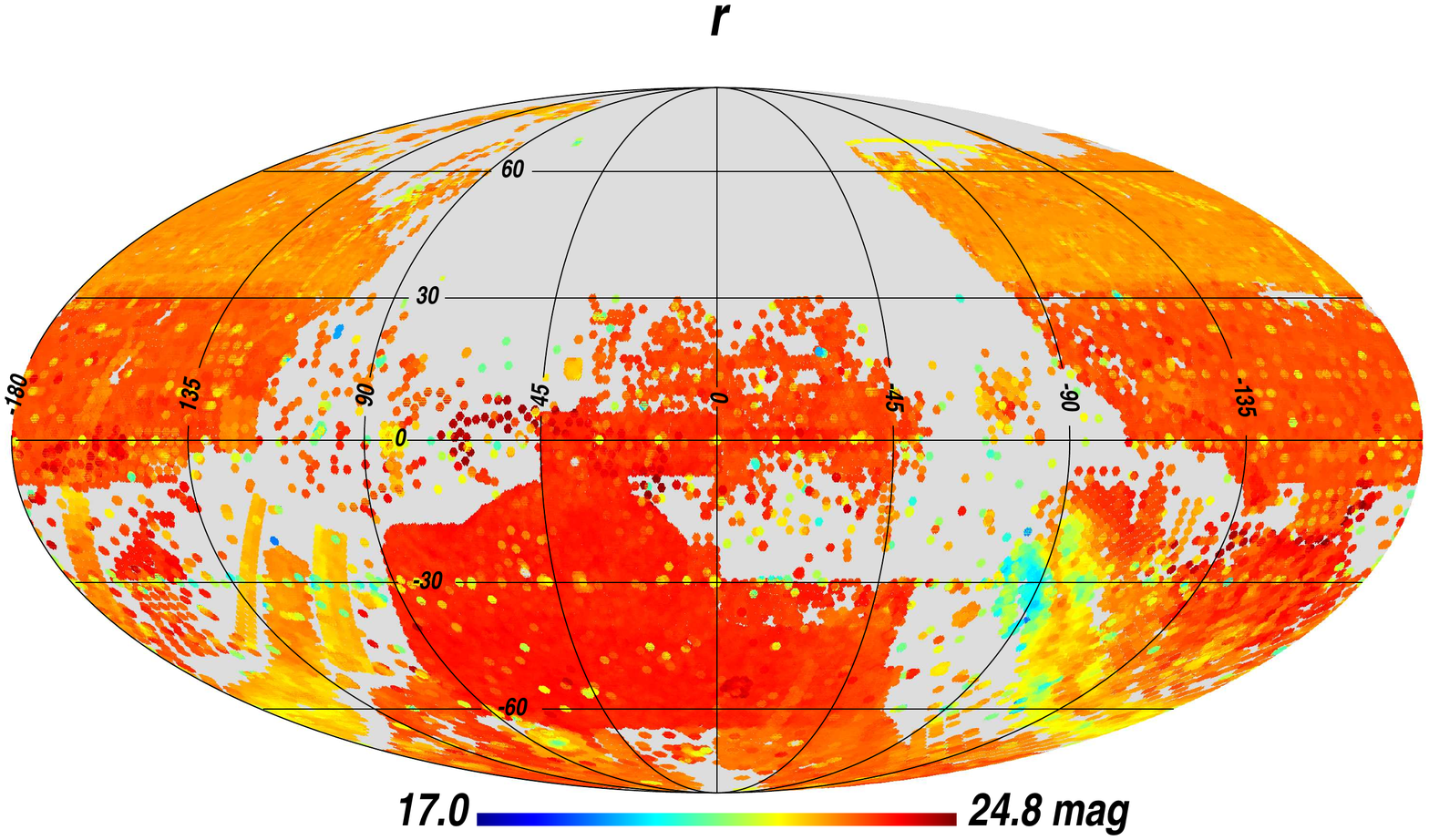}
\includegraphics[trim={0cm 5cm 2cm 1cm},clip,width=0.50\hsize,angle=0]{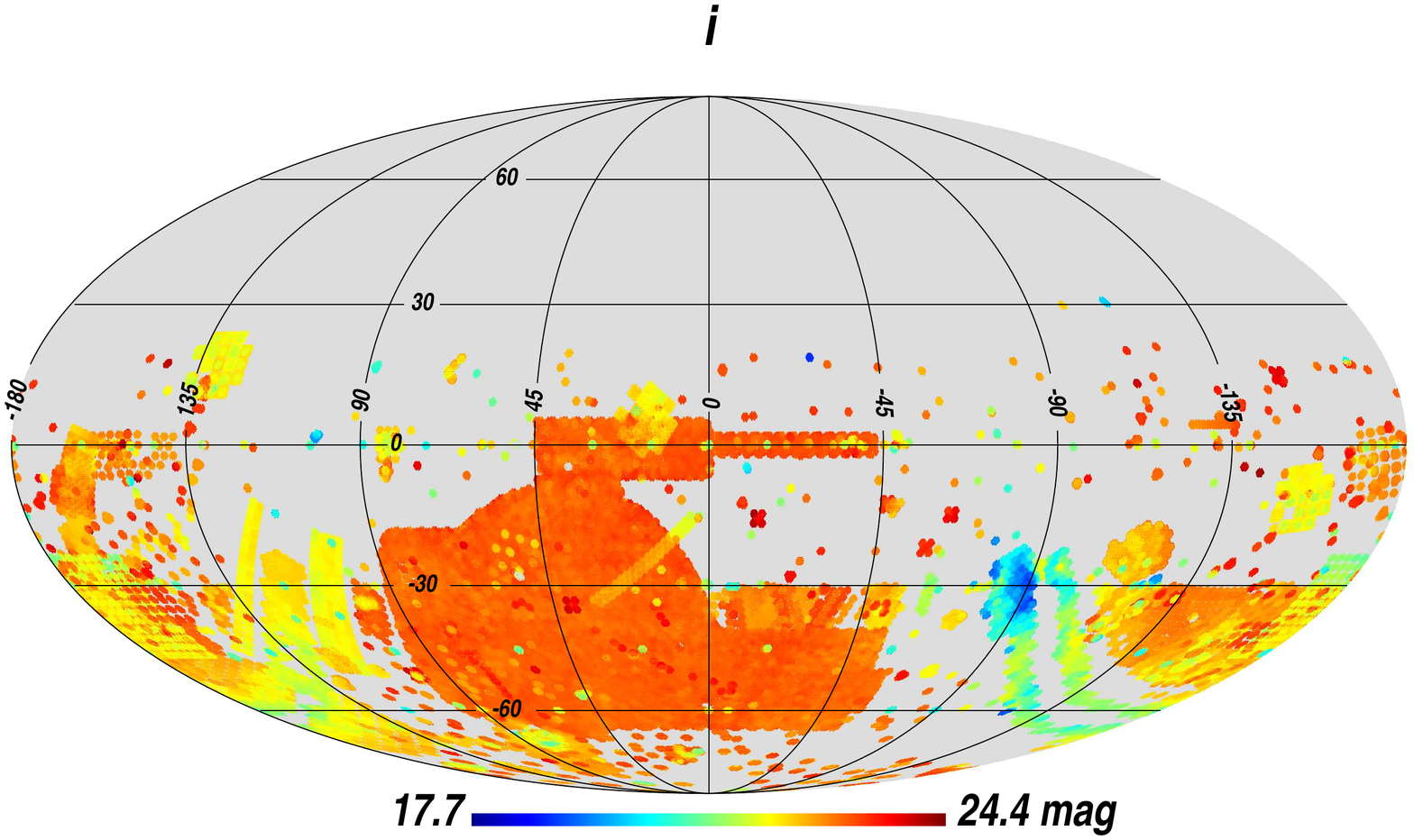} \\
\includegraphics[trim={0cm 5cm 2cm 1cm},clip,width=0.50\hsize,angle=0]{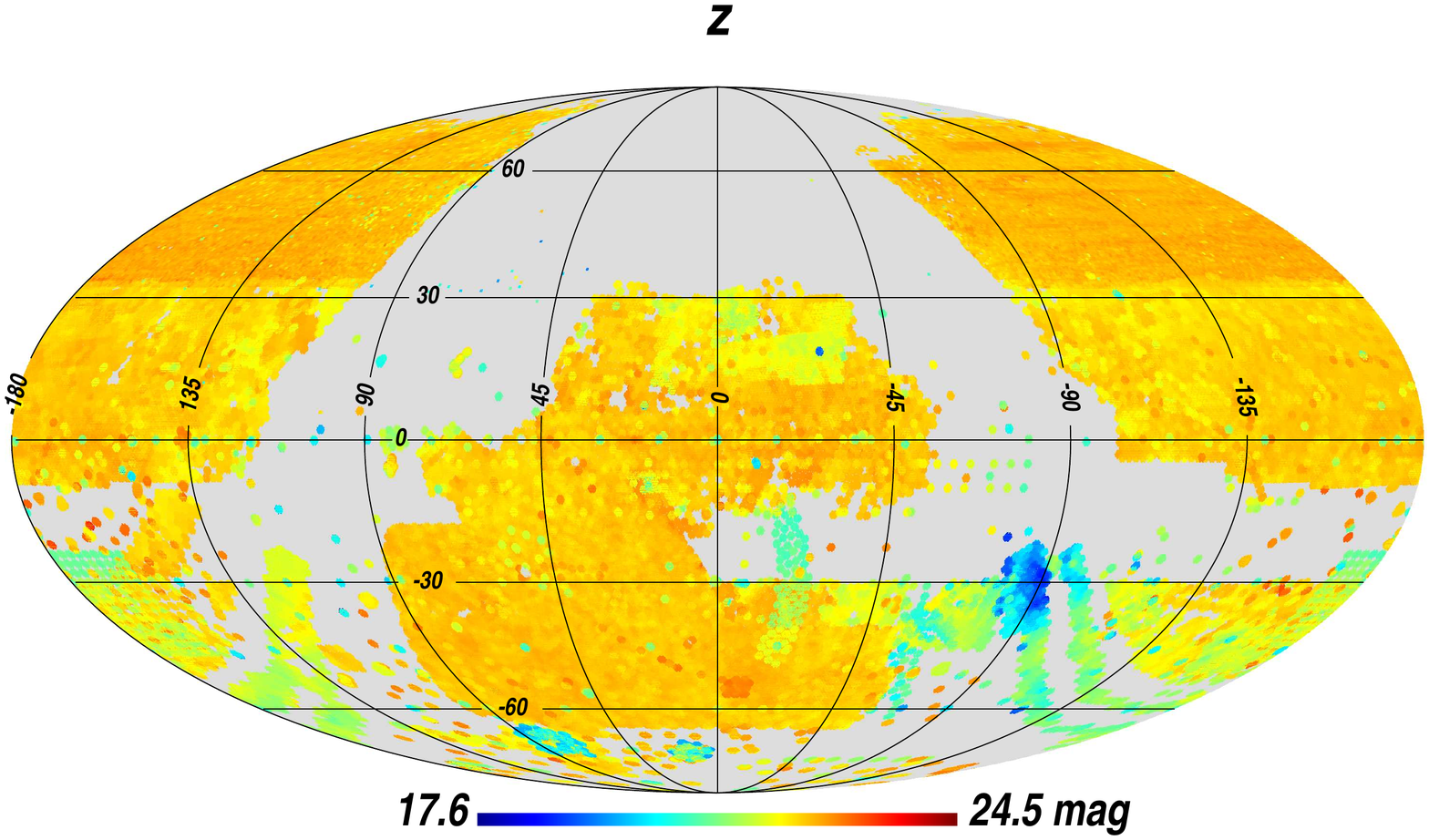}
\includegraphics[trim={0cm 5cm 2cm 1cm},clip,width=0.50\hsize,angle=0]{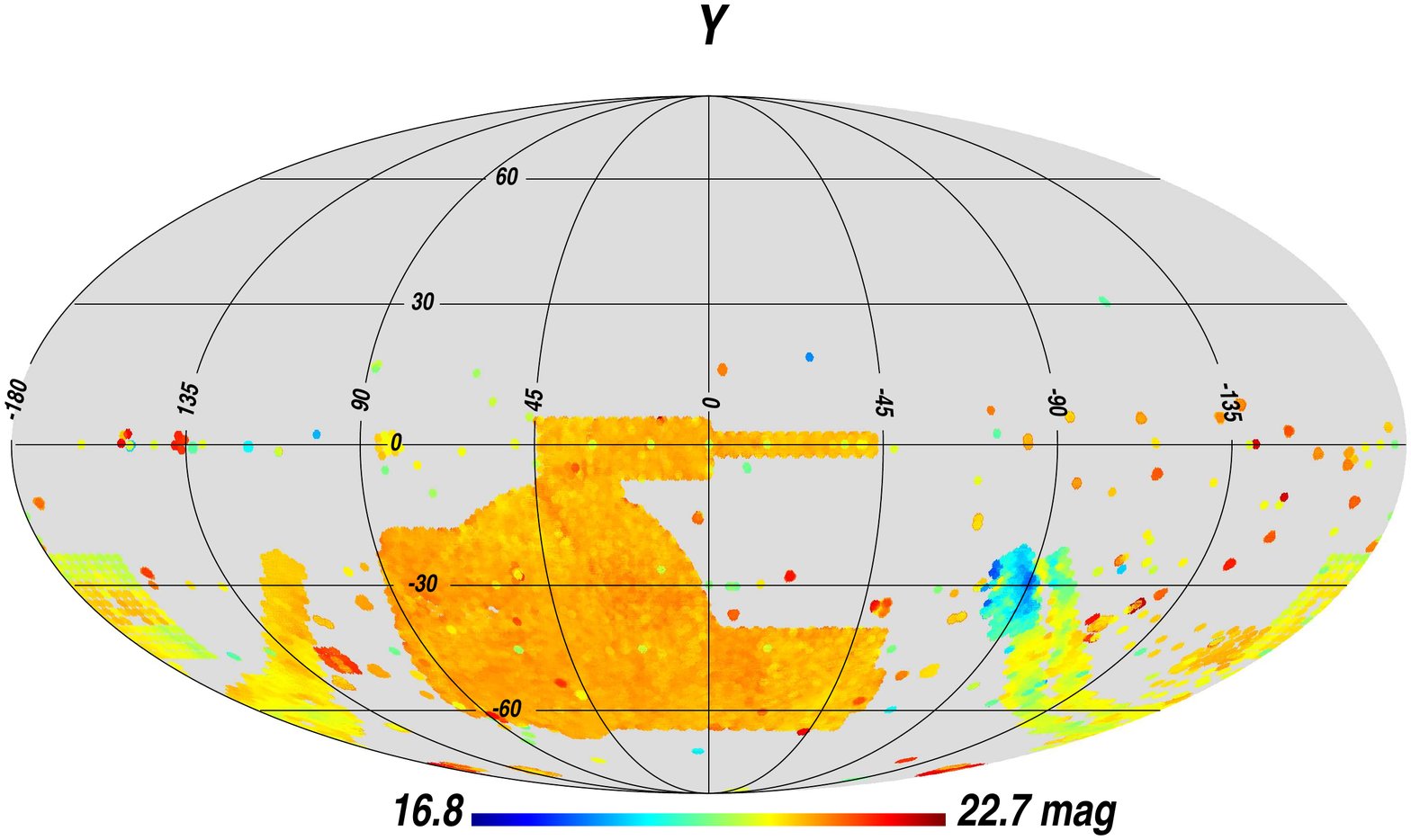} \\
\includegraphics[trim={0cm 5cm 2cm 1cm},clip,width=0.50\hsize,angle=0]{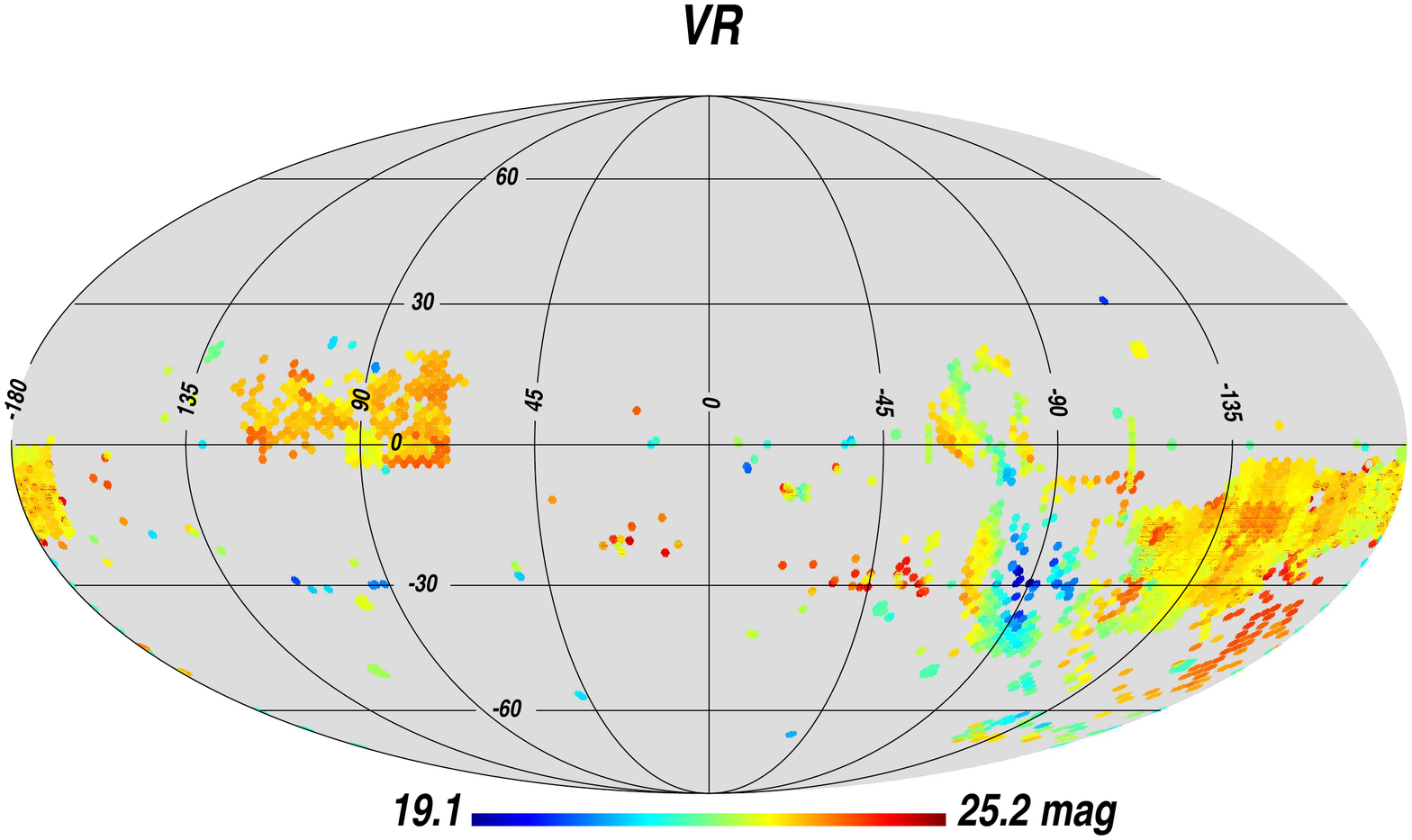}
\end{array}$
\caption{Depth maps (95th percentile) for all seven $u, g, r, i, z, Y$ and {\em VR} bands in equatorial Aitoff projection.}
\label{fig_depths}
\end{figure*}
\end{center}

\begin{center}
\begin{figure*}[ht]
\includegraphics[trim={0cm 5cm 2cm 1cm},clip,width=1.0\hsize,angle=0]{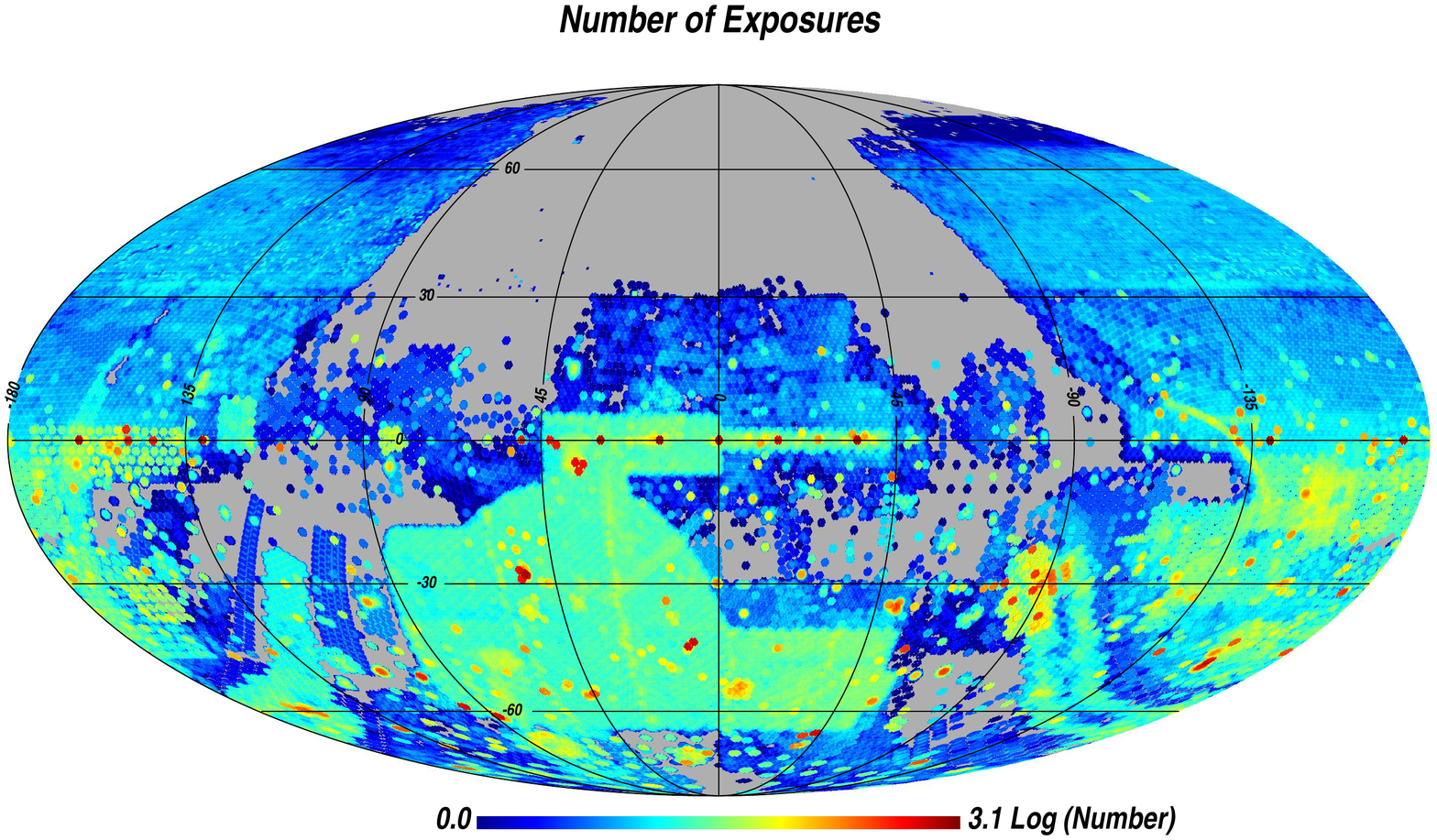}
\caption{Number of NSC exposures on a logarithmic scale in equatorial coordinates.}
\label{fig_nexp}
\end{figure*}
\end{center}

\begin{center}
\begin{figure*}[ht]
\includegraphics[width=1.0\hsize,angle=0]{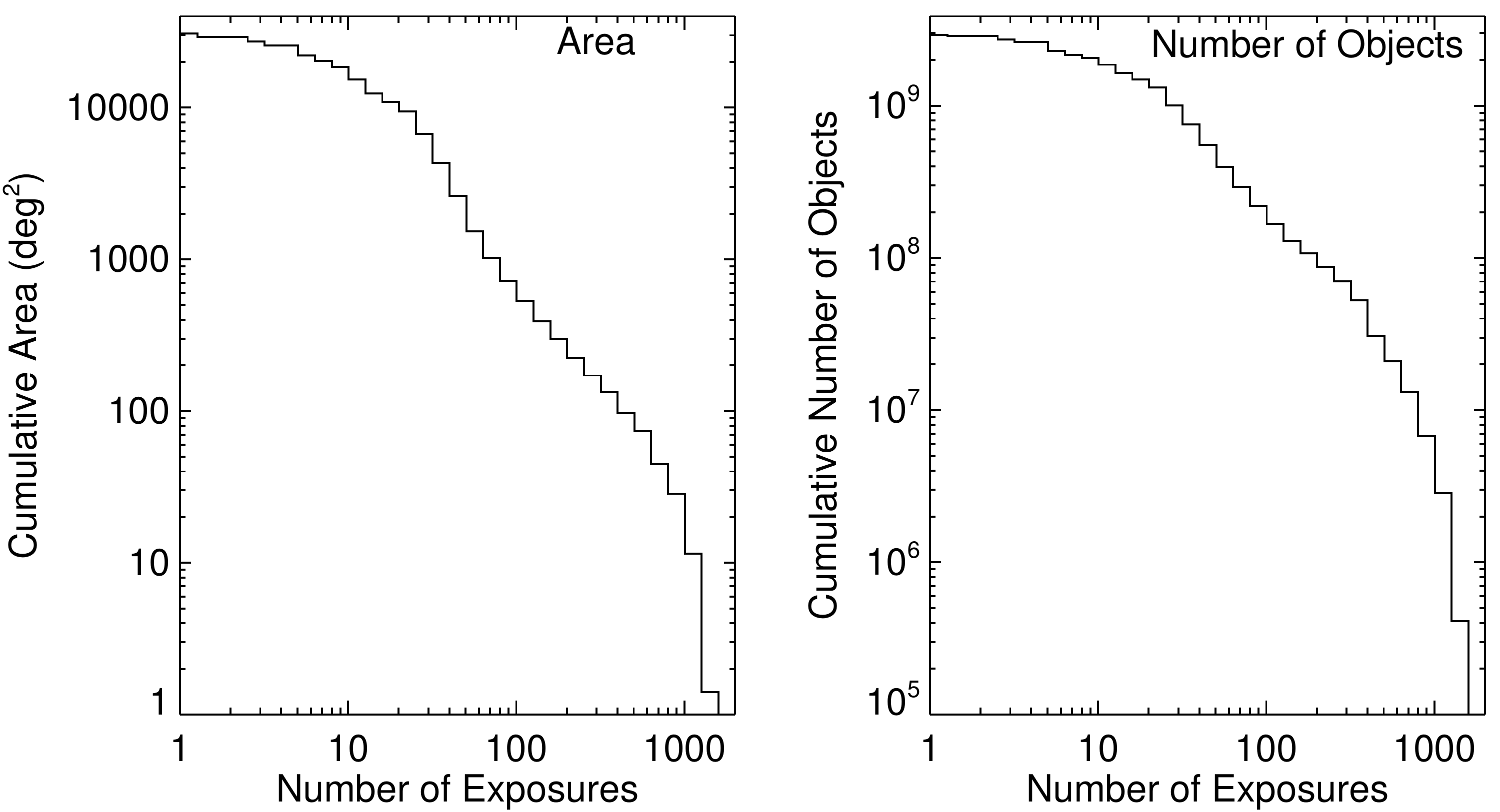}
\caption{Cumulative histogram of ({\em Left}) area and ({\em Right}) number of objects with numbers of exposures greater than some value.}
\label{fig_nexp_cumhist}
\end{figure*}
\end{center}

Figure \ref{fig_bigmap} shows the density of the 2.9 billion objects in the NSC DR1 catalog covering $\approx$30,000 square degrees. There are 34 billion measurements from 255,454 exposures.  Figure \ref{fig_depths} shows maps of the 95th percentile depths in each of the seven bands.  The median depth is
23.1, 23.3, 23.2, 22.9, 22.2, 21.0, 23.1 mag in the $u, g, r, i, z, Y,$ and {\em VR} bands, respectively.
The photometric RMS of bright ($\lesssim$17--18 mag), unsaturated stars in regions with many exposures (e.g., Figure \ref{fig_photscatter_mag}) is $\lesssim$1\% in all bands except for $u$-band where it is $\sim$2\% indicating the precision of the NSC photometry across most of the sky.
The photometric RMS is higher in the Milky Way midplane and the central regions of the Magellanic Clouds likely due to crowding effects.
There are systematic offsets in the regions of high crowding and extinction such as the Milky Way midplane and we advise caution when using the data in these regions as there might be some offsets in the photometric zero points.
Figure \ref{fig_nexp} shows a map of the number of exposures and Figure \ref{fig_nexp_cumhist} shows a cumulative histogram of the number of objects and area with a certain number of exposures or greater.  A significant area on the sky and number of objects have multiple measurements that are useful for time-series science (see Figure \ref{fig_nexp}).  Finally, $\approx$2 billion objects and $\approx$21,000 square degrees of sky have photometry in three or more bands that are usable for constructing color-color diagrams.

The NSC data are released through the NOAO Data Lab\footnote{\url{https://datalab.noao.edu}} \citep{Fitzpatrick2016}.
Access and exploration tools include the Data Lab Data Discovery tool, database access to the catalog (via direct SQL query or TAP service), an image cutout service, and a Jupyter notebook server with example notebooks for exploratory analysis. 

\begin{center}
\begin{figure*}[t]
$\begin{array}{cc}
\includegraphics[width=0.5\hsize,angle=0]{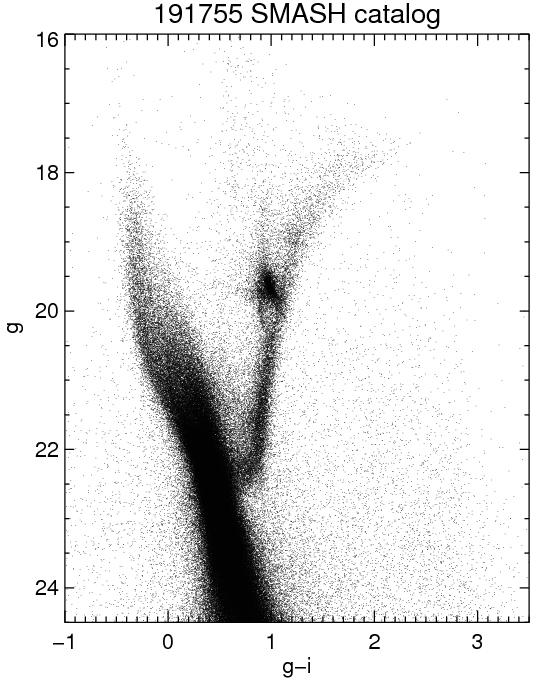}
\includegraphics[width=0.5\hsize,angle=0]{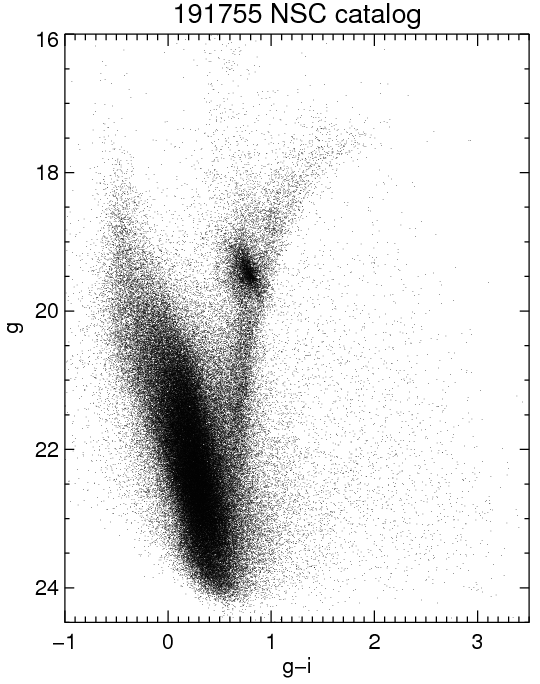}
\end{array}$
\caption{CMD comparison for HEALPix pixel 191755 at a radius of 2.4\dgr from the LMC center. {\em (Left)} SMASH photometry.
{\em (Right)} NSC photometry.}
\label{fig_cmdcomparison}
\end{figure*}
\end{center}

\begin{center}
\begin{figure}[t]
\includegraphics[width=1.0\hsize,angle=0]{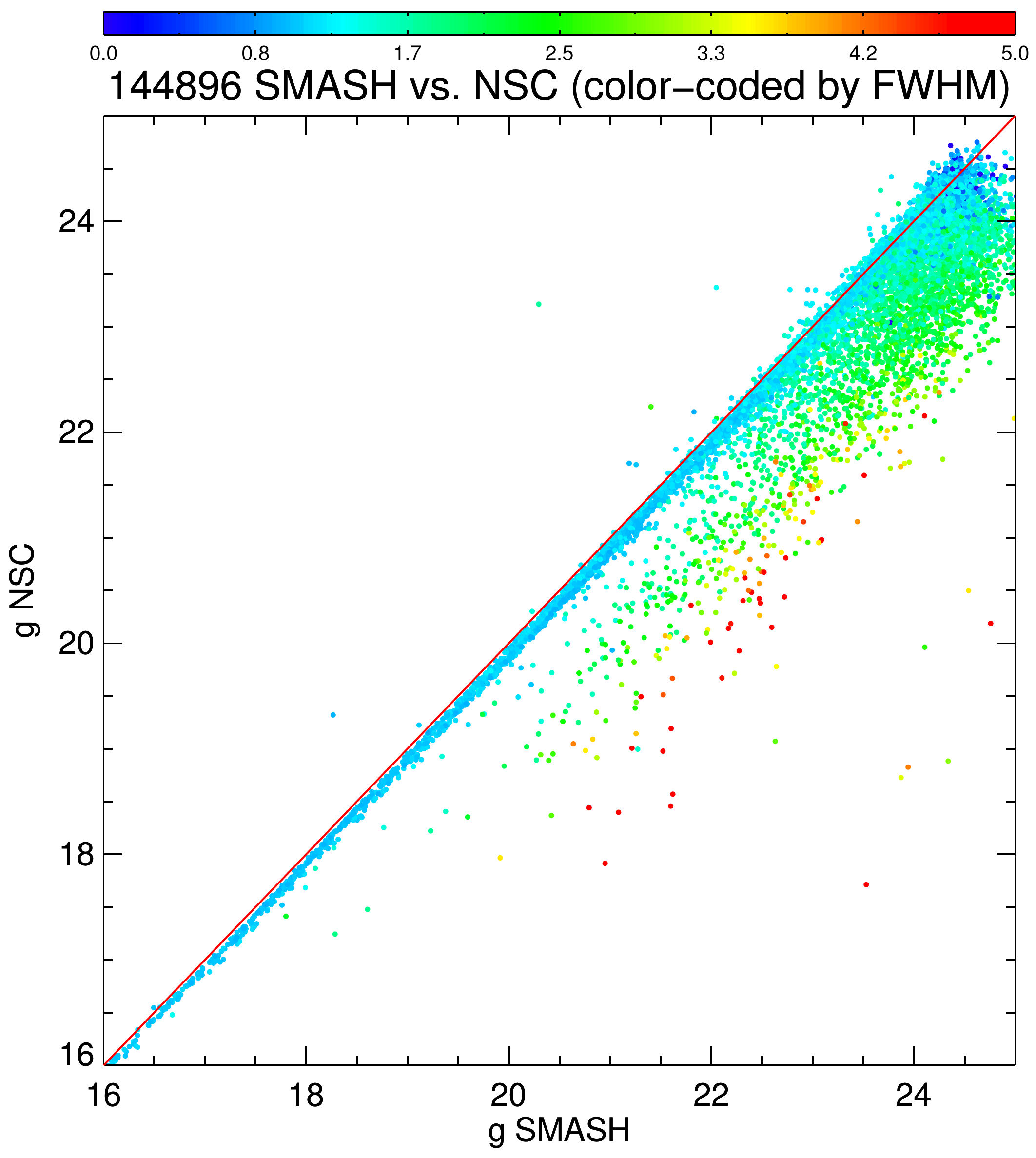}
\caption{Direct $g$-band one-to-one comparison of SMASH and NSC photometry in the HEALPix pixel 144896 color-coded by an object's FWHM.  The one-to-one line is shown in red.}
\label{fig_photcompare}
\end{figure}
\end{center}

Figure \ref{fig_cmdcomparison} shows a comparison of the NSC photometry with the SMASH \citep{Nidever2017} photometry for a HEALPix pixel in a crowded field only 2.4\dgr from the LMC center.  The SMASH CMD is deeper because it uses forced PSF photometry with detection on a multi-band stacked image and the stellar population features are sharper due to the lower photometric uncertainties from the PSF photometry.  However, most of the important features are still visible in the CMD using the NSC aperture photometry in this crowded field.  A direct $g$-band one-to-one comparison of crossmatched SMASH and NSC objects in the HEALPix pixel 144896 is shown in Figure \ref{fig_photcompare}.  The photometry compares well for sources with low FWHM (i.e., point sources; shown in light blue) while they deviate for extended sources as expected because SMASH only has PSF photometry.  Even for point sources there is a slight zero point magnitude offset ($\approx$0.1 mag) because SMASH and NSC are on slightly different photometric systems.

\begin{center}
\begin{figure}[t]
\includegraphics[width=1.0\hsize,angle=0]{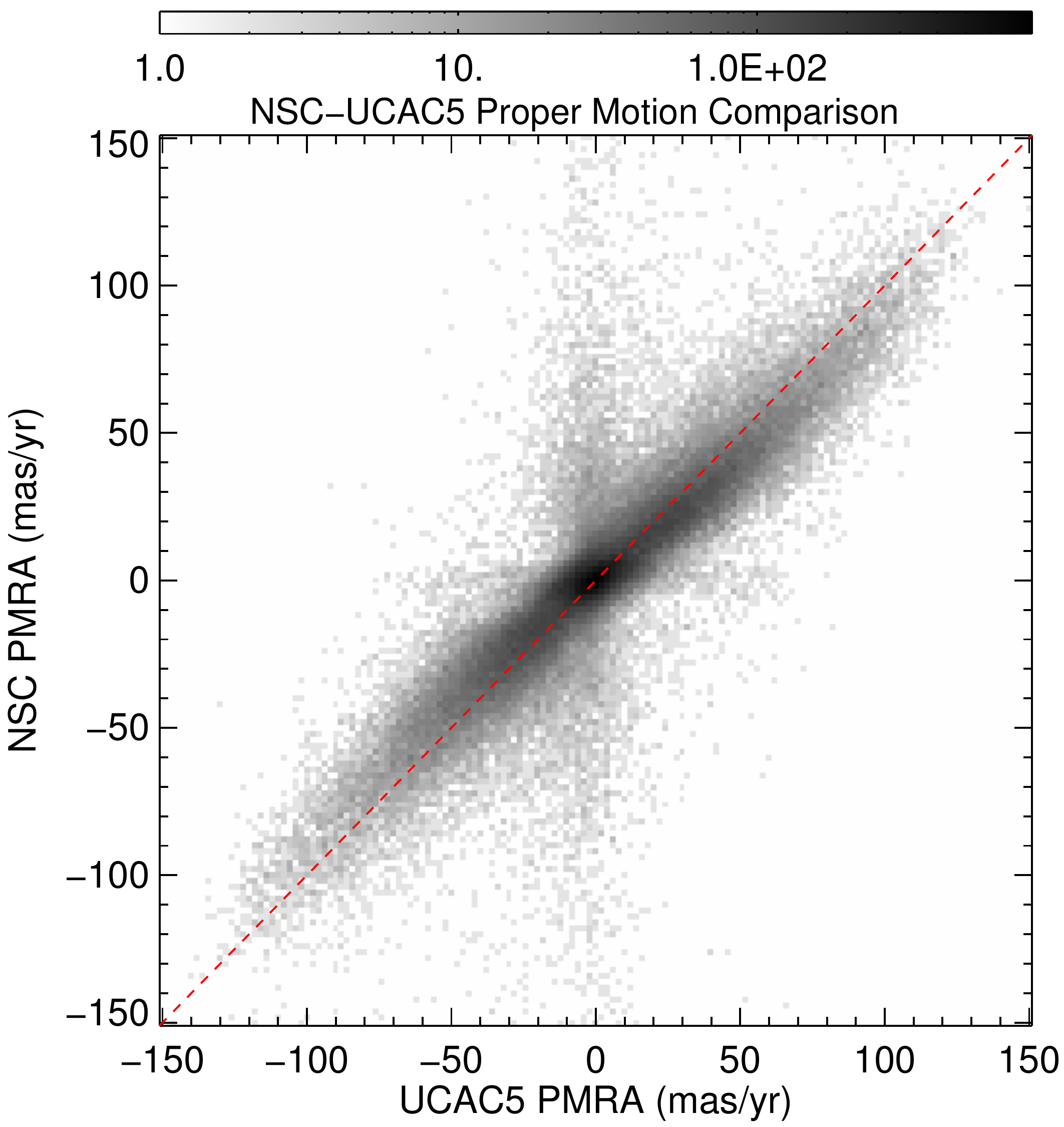}
\caption{Comparison of NSC proper motions to those from UCAC5. The S/N of the proper motion measurement was required to be higher than 3 or the proper motion error less than 3 mas yr$^{-1}$ in both catalogs.  The one-to-one line is shown in red.}
\label{fig_pmcomparison}
\end{figure}
\end{center}

Figure \ref{fig_pmcomparison} shows an all-sky comparison of well-measured NSC proper motions to those in UCAC5 \citep{Zacharias2017}.  The two datasets compare well with a small number of outliers at low UCAC5 proper motion values but higher NSC values.

% proper motion comparison
% morphology comparison

% number of cpu-hours used

There are six types of NSC DR1 tables in the database:
\begin{itemize}
\item {\bf Coverage}. Coverage information for each nside=4096 HEALPix pixel with a resolution of 0.86\arcmin.  This gives the coverage fraction and depth for each band.  201,326,592 rows.
\item {\bf Chip}. Information for each unique chip image with astrometric correction terms and uncertainties. 12,008,634 rows.
\item {\bf Exposure}. Information for each unique exposure with zero-point values and uncertainties. 255,454 rows.
\item {\bf Object}. Information on each unique object with average photometric, astrometric and morphology values. 2,930,644,736 rows.
\item {\bf Measurement}. Information on each individual source measurement in single-epoch images. 34,658,213,888 rows.
\item {\bf Crossmatch}. Crossmatch information between the NSC and the DESI Legacy Surveys, ALLWISE, and DES.
\end{itemize}

\section{Example Science Use Cases}
\label{sec:usecases}

Below, we summarize a few example science use cases of the NSC; there are clearly many other possible science applications.

\subsection{Dwarf Galaxies}
\label{subsec:dwarfgalaxies}

The Milky Way is orbited by faint dwarf galaxy satellites that can be challenging to detect.
The best method to discover these objects has been with resolved stellar populations
as recently demonstrated by \citet{Bechtol2015} and \citet{Drlica-Wagner2015} using DECam data from the Dark Energy Survey.  The NSC includes data of resolved stellar populations for regions of the sky previously unexplored for dwarf galaxies.  The NOAO Data Lab has an example Jupyter notebook illustrating how to search for dwarf galaxies in the NSC.

\subsection{Stellar Streams}
\label{subsec:stellarstreams}
% help from Knut?

The Milky Way halo hosts a number of stellar streams from tidally stripped galaxies and globular clusters.  One of the most famous streams is the Sagittarius stellar stream \citep[e.g.,][]{Majewski2003,Koposov2012} that stretches more than 360\degr~around the sky.  Under the currently-favored hierarchical galaxy formation model, galaxies like the Milky Way are formed through the continual merging and accretion of smaller galaxies. The streams that we currently see are relics of this process.  Studying these streams can help us understand how our Galaxy was formed and its accretion history. The NSC covers regions of the sky that have not yet been explored for stellar streams.
% help from Knut?

\subsection{The Variable Sky}
\label{subsec:variables}

Many astronomical objects vary with time either in their brightness or position in the sky.  The temporal information in the NSC can be used to investigate the photometric and astrometric variability of objects.  The NSC has 10,000 square degrees with $\sim$20 exposures, 200 square degrees with $\sim$100 exposures and 10 square degrees with at $\sim$1000 exposures (also see Figures \ref{fig_nexp} and \ref{fig_nexp_cumhist}). These data, combined with priors based on population statistics \citep{Narayan}, can provide probabilistic classifications for a large fraction of known variable types.
\begin{itemize}
\item Variable stars, such as RR Lyrae, are essentially standard candles and are used to explore stellar structures in the Milky Way halo. The NSC sensitivity to RR Lyrae stars extends throughout the Milky Way and its satellites.
\item Active Galactic Nuclei (AGN) host supermassive black holes at the centers of galaxies and investigating their photometric variability can be used to study the accretion history of material onto the black hole, estimate the black hole mass, and explore the structure of the broad emission-line region.
% help from Stephanie and Robert?
\item Transient events, like supernova explosions can be used to study interesting astrophysical phenomena as well as the expansion of the universe. NSC images document the progenitor star fields of future SNe.
\item Proper motions of objects can be used to identify moving groups of stars in the Milky Way galaxy and to select nearby, low luminosity objects by their large apparent motion.
\end{itemize}
The NSC can be used to study all of these phenomena across nearly the entire sky.
% help from Steve

% Steve's text about NSC usefulness to variability studies
NSC provides 1000 images for $\sim$2$\times$10$^6$ targets, and 100 images for $\sim$2$\times$10$^8$ targets.  This is sufficient to investigate brighter examples of the most numerous object types for which LSST will provide variability details, and to evaluate predictions of target counts.

A simple binary population model \citep{Ridgway1} shows that faint, potentially interacting high-energy binary systems should be numerous, up to $\sim$1000 per deg$^2$ at mid-galactic latitudes.  It is not known what fraction of these may provide detectable outbursts with what frequency, and the NSC can constrain this fraction.

Predictions for the content of the variable sky \citep{Ridgway2} indicate that at the NSC depth, the sky should contain $\sim$100 detectably variable QSOs, and $\sim$500 detectably variable AGN per deg$^2$ -- predictions that can be refined with NSC statistics.

The NOAO Data Lab has a Jupyter notebook showing how to work with light curves for RR Lyrae stars in a catalog like the NSC.

\subsection{Synoptic Sky Surveys}
\label{subsec:synotic}

The NSC is intermediate in depth among the Pan-STARRS \citep{Kaiser2002}, SDSS \citep{York00}, DES \citep{DES} and Megacam \citep{Boulade2003} surveys, and has both partial overlap and complementary coverage with respect to each.

The NSC $5\sigma$ depth of $\approx$23rd magnitude is a good match for the synoptic survey capability of the Zwicky Transient Factory \citep{Bellm2015}, which is projected to have a limiting magnitude $\sim$20.5.  Most detectably variable objects are Galactic stars, and the NSC catalogs a large fraction of those. NSC photometry will provide valuable color and variability diagnostics for common Galactic transient types such as flares from cool dwarfs and outbursts of cataclysmic variables.  In addition, the NSC can be used to provide characteristics of host galaxies for extragalactic variability phenomena such as AGN, QSOs and Tidal Disruption Events.  The NSC can provide historical evidence that candidate SNe are in fact blank field transients, and not outbursts of known objects.  The NSC can also provide supplemental variability information, with a longer temporal baseline, for all observed source types.  
Other surveys such as PS1 and Gaia will also provide similar information for synoptic surveys.

Finally, the NSC will also serve as a first epoch of astrometric and multiple epochs of historical photometric data for the LSST surveys, which will allow for the measurement of accurate proper motions and long-term variability for objects fainter than Gaia's magnitude limit.

\subsection{Solar System Objects}
\label{subsec:solar}

The NSC temporal information can be used to detect solar system objects. Undoubtedly, trans-Neptunian objects detected in the NSC catalog remain to be identified. There is a large interest in finding near earth objects that could potentially impact the earth, and NSC can be used to ``precover" objects that will be discovered in future surveys. For instance, Planet 9 \citep{Batygin2016}\footnote{\url{http://www.findplanetnine.com}}, if it exists, could be lurking inside the NSC.

\section{Future Plans}
\label{sec:future}

%-future plans
%  -PSF photometry of individual exposures
%  -PSF photometry of coadds
%  -forced photometry?
%  -real-time upgrades
%  -possibly catalog-level transient detection/searches
% -real "absolute" proper motions with possibly a better
%   first epoch

Our approach with NSC DR1 was to provide a simple and astronomically useful catalog for users. In future data releases, we plan to improve the NSC in the following ways:

%For the first data release of the NSC we took the approach to start very simple and once we understood the main steps we would increase the complexity of the analysis.  Here are some of the improvements we plan to make in the future:

\begin{itemize}
\item Improved photometric calibration in the south: SkyMapper data \citep{Wolf2018} will be used for photometric calibration and should improve the photometric accuracy.  This should also allow for chip-level zero points.
\item Improved proper motions: The current proper motions are relative to the foreground stars and limited to $\lesssim$200 mas yr$^{-1}$ because of the 0.5\arcsec~matching radius used in the combination step.  In the future, the proper motions will be put on an absolute scale by incorporating Gaia proper motions and parallaxes in our astrometric calibration, applying position-dependent correction terms, or using an extragalactic astrometric reference frame (e.g., QSOs). We will also explore using improved cross-matching algorithms to detect fast-moving objects.
\item Improved photometric variability index: Better photometric variability quantities will be computed including the $\chi^2$ and probability value of the object being photometrically constant and possibly Lomb-Scargle \citep{Lomb1976,Scargle1982} like metrics \citep[e.g.,][]{Saha2017}.  In addition, a catalog of variables will be compiled.
\item PSF photometry of individual exposures:  PSF photometry will be performed on the individual images.
\item PSF photometry of coadd images:  Coadd images will be created for each band and PSF photometry performed on them.
\item Forced photometry:  Once the coadd photometry catalogs exist, it will be possible to perform forced photometry on the individual images.
\item Real-time updates: Instead of regular data releases on an annual or bi-annual timescale, the catalog will be updated in real-time as new exposures become public.
\item Catalog-level transient detection:  Transients can be detected at the catalog level if the photometry can be produced fast enough ($<$1 day).
\end{itemize}

\section{Summary}
\label{sec:summary}

We present the first public data release (DR1) of the NOAO Source Catalog (NSC) based on the majority of the public imaging data in the NOAO Science Archive (as of 2017-10-11) from both the northern and southern hemispheres.  The NSC covers $\approx$30,000 square degrees and contains over 2.9 billion unique objects with 34 billion individual measurements to a depth of $\approx$23rd magnitude.  With a baseline of $\approx$5 years, tens to thousands of exposures per object, and an astrometric zero point accuracy of $\approx$7 mas, the NSC provides precise proper motions for many objects.  The NSC will be useful for exploring Galactic structure by searching for dwarf satellite galaxies, stellar streams and mapping variable stars.  In addition, the temporal component of the NSC will allow for interesting time-series studies including searches for transients, investigations of AGN variability, and searches for solar system objects.  The NSC will also serve as a good astrometric and photometric first epoch for future synoptic surveys like ZTF and LSST. Future improvements to the NSC include PSF photometry of individual and stacked images as well as real-time updates to the catalog as new images become public.  We invite users to contribute to our effort and create useful data products that can improve the NSC and become part of DR2.

\acknowledgments

This project used data obtained with the Dark Energy Camera (DECam) at the Blanco 4m telescope at Cerro Tololo Inter-American Observatory. DECam was constructed by the Dark Energy Survey (DES) collaborating institutions: Argonne National Lab, University of California Santa Cruz, University of Cambridge, Centro de Investigaciones Energeticas, Medioambientales y Tecnologicas-Madrid, University of Chicago, University College London, DES-Brazil consortium, University of Edinburgh, ETH-Zurich, University of Illinois at Urbana-Champaign, Institut de Ciencies de l'Espai, Institut de Fisica d'Altes Energies, Lawrence Berkeley National Lab, Ludwig-Maximilians Universit\"at, University of Michigan, National Optical Astronomy Observatory, University of Nottingham, Ohio State University, University of Pennsylvania, University of Portsmouth, SLAC National Lab, Stanford University, University of Sussex, and Texas A\&M University. Funding for DES, including DECam, has been provided by the U.S. Department of Energy, National Science Foundation, Ministry of Education and Science (Spain), Science and Technology Facilities Council (UK), Higher Education Funding Council (England), National Center for Supercomputing Applications, Kavli Institute for Cosmological Physics, Financiadora de Estudos e Projetos, Funda\c{c}\~ao Carlos Chagas Filho de Amparo a Pesquisa, Conselho Nacional de Desenvolvimento Cientfico e Tecnol\'ogico and the Minist\'erio da Ci\^encia e Tecnologia (Brazil), the German Research Foundation-sponsored cluster of excellence "Origin and Structure of the Universe" and the DES collaborating institutions. The Cerro Tololo Inter-American Observatory, National Optical Astronomy Observatory is operated by the Association of Universities for Research in Astronomy (AURA) under a cooperative agreement with the National Science Foundation. 

This project also incorporates observations obtained at Kitt Peak National Observatory, National Optical Astronomy Observatory, which is operated by the Association of Universities for Research in Astronomy (AURA) under cooperative agreement with the National Science Foundation. The paper also contains data from the Bok 90" telescope of Steward Observatory, which is located on Kitt Peak and operated by the University of Arizona. The authors are honored to be permitted to conduct astronomical research on Iolkam Du'ag (Kitt Peak), a mountain with particular significance to the Tohono O'odham. 

This research uses services or data provided by the NOAO Data Lab. NOAO is operated by the Association of Universities for Research in Astronomy (AURA), Inc. under a cooperative agreement with the National Science Foundation.

This publication makes use of data from the Pan-STARRS1 Surveys (PS1) and the PS1 public science archive, which have been made possible through contributions by the Institute for Astronomy, the University of Hawaii, the Pan-STARRS Project Office, the Max-Planck Society and its participating institutes, the Max Planck Institute for Astronomy, Heidelberg and the Max Planck Institute for Extraterrestrial Physics, Garching, The Johns Hopkins University, Durham University, the University of Edinburgh, the Queen's University Belfast, the Harvard-Smithsonian Center for Astrophysics, the Las Cumbres Observatory Global Telescope Network Incorporated, the National Central University of Taiwan, the Space Telescope Science Institute, the National Aeronautics and Space Administration under Grant No. NNX08AR22G issued through the Planetary Science Division of the NASA Science Mission Directorate, the National Science Foundation Grant No. AST-1238877, the University of Maryland, Eotvos Lorand University (ELTE), the Los Alamos National Laboratory, and the Gordon and Betty Moore Foundation.

This work has made use of data from the European Space Agency (ESA)
mission {\it Gaia} (\url{https://www.cosmos.esa.int/gaia}), processed by
the {\it Gaia} Data Processing and Analysis Consortium (DPAC,
\url{https://www.cosmos.esa.int/web/gaia/dpac/consortium}). Funding
for the DPAC has been provided by national institutions, in particular
the institutions participating in the {\it Gaia} Multilateral Agreement.

This publication makes use of data products from the Two Micron All Sky Survey, which is a joint project of the University of Massachusetts and the Infrared Processing and Analysis Center/California Institute of Technology, funded by the National Aeronautics and Space Administration and the National Science Foundation.

We acknowledge with thanks the variable star observations from the AAVSO International Database contributed by observers worldwide and used in this research.

%\bibliography{/Users/martin/Work/Papers/Biblio}
%\bibliographystyle{apj}

% Bibtex will create a .bbs file in the directory and before sending to the editor, I should replace the bibliography call by this file.

\end{document}